# The sharp turn: backward rupture branching during the 2023 $M_\mathrm{w}$ 7.8 Turkey earthquake

Xiaotian Ding[1], Shiqing Xu[1*], Yuqing Xie[2], Martijn van den Ende[2], Jan Premus[2], and Jean-Paul Ampuero[2]

[1]Department of Earth and Space Sciences, Southern University of Science and Technology, Shenzhen, China

[2]Université Côte d'Azur, IRD, CNRS, Observatoire de la Côte d'Azur, Geoazur, Nice, France

[*]Corresponding author: Shiqing Xu (xusq3@sustech.edu.cn), https://orcid.org/0000-0001-5968-1320

## Key points

1. Rupture started on a splay fault, then branched bilaterally on the East Anatolian Fault (EAF), including onto a fault segment at acute angle to the splay fault.
2. Dynamic rupture models explain the mainshock rupture path, including its apparent backward branching, and provide insight on its controlling factors.
3. Rupture on the SW EAF segment can be triggered by rupture on the NE EAF segment or, if the EAF is highly pre-stressed, by the initial splay-fault rupture.

## Abstract

Multiple lines of evidence indicate that the 2023 $M_\mathrm{w}$ 7.8 Turkey earthquake started on a splay fault, then branched bilaterally onto the nearby East Anatolian Fault (EAF). This rupture pattern includes one feature deemed implausible, called backward rupture branching: rupture propagating from the splay fault onto the SW EAF segment through a sharp corner (with an acute angle between the two faults). To understand this feature, we perform 2.5-D dynamic rupture simulations considering a large set of possible scenarios. We find that both subshear and supershear ruptures on the splay fault can trigger bilateral ruptures on the EAF, which themselves can be either subshear, supershear, or a mixture of the two. In most cases, rupture on the SW segment of the EAF starts after rupture onset on its NE segment: the SW rupture is triggered by the NE rupture. Only when the EAF has initial stresses very close to failure, its SW segment can be directly triggered by the initial splay-fault rupture, earlier than the activation of the NE segment. These results advance our understanding of the mechanisms of multi-segment rupture and the complexity of rupture processes, paving the way for a more accurate assessment of earthquake hazards.

## Plain language summary

The 2023 $M_w$ 7.8 Turkey earthquake ruptured multiple fault segments, featuring an unexpected backward rupture branching through a sharp corner: rupture initially propagated toward northeast on a splay fault and then made a nearly U-turn onto the southwest segment of the nearby East Anatolian Fault (EAF). To understand such intriguing feature, we conduct a series of computer simulations of earthquake ruptures. Our results show that the sharp-turn rupture can be realized in two different ways: (i) rupture first jumps ahead from the splay fault to the northeast segment of the EAF, which later triggers rupture on the southwest segment the EAF, and (ii) the initial splay-fault rupture directly triggers rupture on the southwest segment of the EAF. The realization of (i) or (ii) depends on the initial stress and friction conditions, and hence provides useful clues for understanding the preconditions and detailed rupture process of the 2023 $M_w$ 7.8 earthquake. The results also shed light on anticipating possible rupture paths and maximum earthquake magnitude in other regions containing multiple fault segments.

## 1. Introduction

Multiple fault segments can rupture during a single earthquake or earthquake sequence. Examples include the 1992 $M_w$ 7.3 Landers earthquake in the Eastern California Shear Zone (ECSZ) (Sieh et al., 1993), the 2010 $M_w$ 7.2 El Mayor-Cucapah earthquake in Baja California (Wei et al., 2011), the 2012 $M_w$ 8.6 earthquake off the coast of Sumatra (Meng et al., 2012a; Yue et al., 2012), the 2016 $M_w$ 7.8 Kaikōura earthquake in New Zealand (Hamling et al., 2017; Wang et al., 2018), and the 2019 Ridgecrest earthquake sequence to the north of the ECSZ (Ross et al., 2019). Theoretical, numerical and laboratory studies have been conducted to understand how and why multi-segment ruptures can occur (DeDontney et al., 2012; Duan and Oglesby, 2007; Harris and Day, 1993; Kame et al., 2003; Poliakov et al., 2002; Rousseau and Rosakis, 2009) and to reproduce the patterns of observed multi-segment earthquakes in dynamic rupture simulations (Wollherr et al, 2019; Ulrich et al., 2019).

In addition to stimulating scientific investigations on earthquake physics, the occurrence of multi-segment ruptures is a challenge for earthquake hazard assessment: how to estimate the maximum magnitude of earthquakes in a region that contains multiple fault segments? Rules of thumb have been proposed based on past earthquake observations, geometrical parameters of faults, and dynamic rupture theory and simulations (Biasi and Wesnousky, 2021; Bohnhoff et al., 2016; Mignan et al., 2015; Walsh et al., 2023). One scenario that has been deemed implausible is rupture branching through a sharp corner characterized by an acute angle (<90 degrees) between the two faults—called backward rupture branching. The main reasons for discarding this scenario are that earlier dynamic rupture studies only considered rupture branching through a gentle corner associated with an obtuse angle between the two faults—called forward rupture branching (Poliakov et al., 2002; Kame et al., 2003), and that backward rupture branching was generally thought to be inhibited by the shear stress release (stress shadow effect) on the backward quadrants induced by the first fault rupture. Nonetheless, backward rupture branching has been observed during some strike-slip earthquakes, with the same or opposite sense(s) of slip along different fault segments (Fliss et al., 2005; Li et al., 2020; Oglesby et al., 2003). Backward rupture branching can occur in subduction zones as well, with thrust or mixed thrust/normal faulting mechanism(s) along different fault segments (Cubas et al., 2013; Melnick et al., 2012; Wendt et al., 2009; Xu et al., 2015). All these new results challenge the simple consideration of the stress shadow effect, and raise questions about the possible rupture paths during large earthquakes, in particular how and why backward rupture branching may occur.

The February 6, 2023 $M_w$ 7.8 Turkey earthquake, while devastating (Dal Zilio and Ampuero, 2023; Hussain et al., 2023), was densely recorded and provides a unique opportunity to address the aforementioned questions related to rupture branching. This earthquake struck in the southwestern stretch of the East Anatolian Fault (EAF) zone, an active plate boundary that accommodates the deformation between the Anatolian plate (AT) and the Arabian plate (AR), dominated by left-lateral shear (Figure 1). The rupture started on a previously unmapped splay fault, about 15 km away from the main EAF strand (Melgar et al., 2023). After arriving at the fault junction, the rupture continued on the Pazarcık segment of the EAF and then propagated bilaterally along the EAF, thus comprising a forward rupture branching to the northeast (NE) and a backward rupture branching to the southwest (SW). The NE-ward rupture finally stopped at around 38°N/38.5°E along the Erkenek segment, while the SW-ward rupture terminated at around 36°N/36°E along the Amanos segment (Figure 1). The total rupture length reached about 350 km and the peak slip 8-12 m (Barbot et al., 2023; Goldberg et al., 2023; Mai et al., 2023; Melgar et al., 2023; Okuwaki et al., 2023). The overall co-seismic slip was dominated by left-lateral strike-slip, along both the splay fault where the earthquake hypocenter was located and the EAF where most of the strain energy was released. This multi-segment rupture came as a surprise, since the most recent large events (with magnitude around and above 7) in the region were confined within individual segments, such as the 1795 $M$ 7.0 earthquake on the Pazarcık segment, the 1872 $M$ 7.2 earthquake on the Amanos segment, and the 1893 $M$ 7.1 earthquake on the Erkenek segment (Güvercin et al., 2022). The multi-segment rupture with an apparent backward rupture branching feature motivates us to investigate the rupture process of the $M_w$ 7.8 earthquake, especially in its early stage.

In Section 2, we characterize the rupture path of the 2023 $M_w$ 7.8 Turkey earthquake using multiple types of observations, and further confirm that it includes an apparent pattern of backward branching. In Section 3, we develop 2.5-D dynamic rupture models (two-dimensional models that account for the finite rupture depth) to understand this rupture pattern, and find that it may be realized in two different modes. In the first mode, the SW segment of the EAF is triggered (possibly with a delay) not by the initial splay-fault rupture but by the rupture on the NE segment of the EAF. The second mode involves early dynamic triggering of the SW segment of the EAF by the initial splay-fault rupture. Our simplified modeling approach focuses on exploring a range of possible scenarios, not on a meticulous comparison between simulated results and observations. Finally, we discuss the implications of the obtained results in Section 4 and draw conclusions in Section 5.

## 2. Mainshock rupture path inferred from observations

Here, we summarize multiple types of observations that help constrain the rupture path during the 2023 $M_w$ 7.8 Turkey earthquake. Although similar results have now been published by various teams, we document here observations that were available immediately after the earthquake (on the same day and up to a few days later) to highlight how rapid seismological products shaped our view of the earthquake rupture and motivated our theoretical work.

### 2.1. Mainshock epicenter, aftershocks and surface rupture trace

The fault geometry is constrained to a first order by the aftershock catalog from the AFAD (Turkey Disaster and Emergency Management Authority, https://deprem.afad.gov.tr) and the surface rupture trace from the USGS (U.S. Geological Survey) (Reitman et al., 2023), as shown in Figure 1. Both an aftershock cluster and a short segment of surface rupture trace delineate a splay fault, where the $M_w$ 7.8 mainshock epicenter was located. Also considering the large offset (15 km) between the relocated earthquake hypocenter and the trace of EAF (Melgar et al., 2023), we can reasonably conclude that the rupture started on the splay fault and then continued on the EAF. The next step is to constrain the rupture branching process, as will be discussed in the following two subsections.

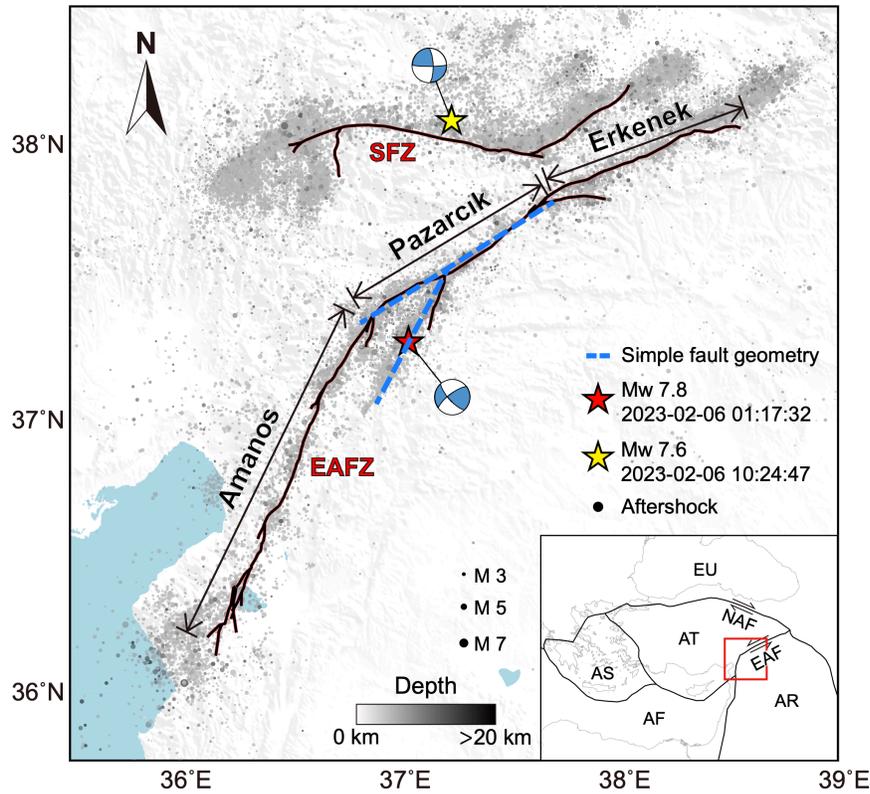

*Figure 1. Distributions of mainshock and aftershock epicenters and surface rupture traces of the 2023 Turkey earthquake sequence. Red and yellow stars show the epicenter location of the $M_w$ 7.8 and $M_w$ 7.6 events, respectively. The beachballs indicate their focal mechanisms determined by the AFAD (Turkey Disaster and Emergency Management Authority) (https://deprem.afad.gov.tr/event-focal-mechanism). The aftershock catalog is also from the AFAD (https://deprem.afad.gov.tr/event-catalog, last accessed on May 29, 2023). Surface rupture traces (black curves with red trimming) are from the USGS (U.S. Geological Survey) (Reitman et al., 2023). EAFZ: East Anatolian Fault Zone. SFZ: Sürgü Fault Zone. From north to south, Erkenek, Pazarcık, and Amanos denote different segments of the EAFZ in the study region (Güvercin et al., 2022). Blue dashed lines depict the simplified fault geometry adopted in our numerical simulations. The inset shows the regional map and tectonic plates. AS: Aegean Sea plate. EU: Eurasian plate. AT: Anatolian plate. AF: African plate. AR: Arabian plate. EAF: East Anatolian Fault. NAF: North Anatolian Fault.*

## 2.2. Teleseismic back-projection

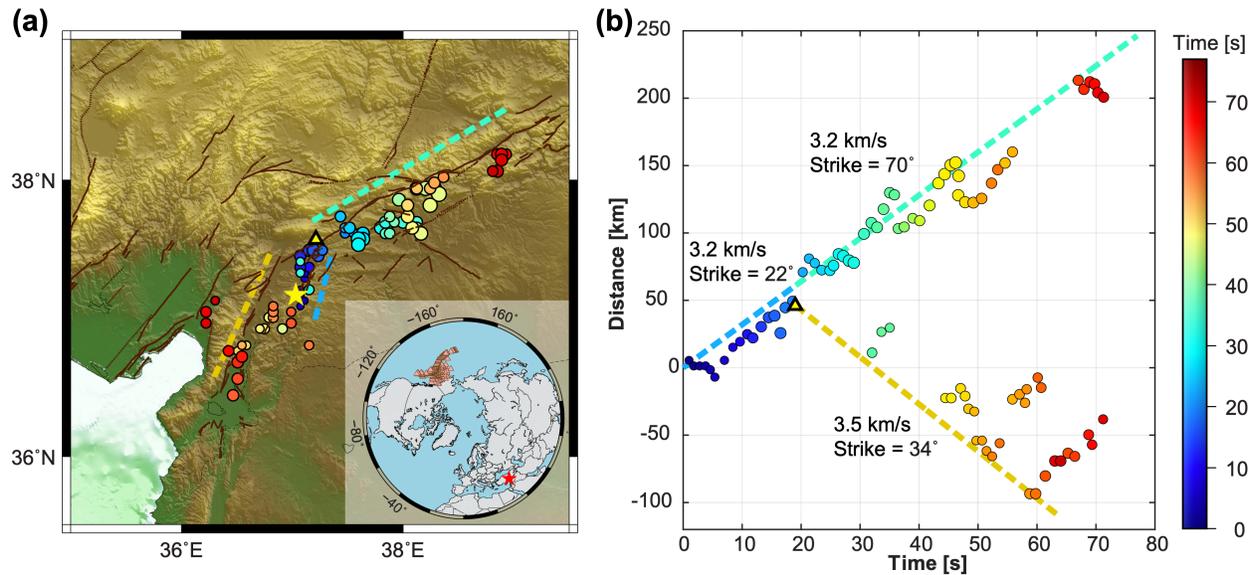

*Figure 2. Teleseismic back-projection imaging of the high-frequency radiation from the M$_w$ 7.8 Turkey earthquake using the Alaska array. (a) Map view of the locations of radiators, color-coded by time and size-coded by power. The yellow star indicates the epicenter reported by the USGS. The brown lines are active faults in Turkey reported by Emre et al. (2013). The inset shows the location of the Alaska array. The pink triangles represent the broadband stations, the red star indicates the epicenter as reported by the USGS. The dashed lines indicate the strike direction of three main fault segments; for clarity, they are offset from the segment traces. (b) Spatiotemporal distribution of radiators. Distance refers to the position along the strike directions shown in (a) as dashed lines; the strike values are shown in (b). The positions of radiators before 19 s are relative to the epicenter, the later ones are relative to the junction between the splay fault and the EAF (37.208ºE, 37.531ºN, according to the surface rupture map from Reitman et al. (2023)), which is indicated by a yellow triangle in (a) and (b). The dashed lines show reference rupture speeds for each segment.*

We image the rupture process by teleseismic back-projection, a method that can image multi-fault ruptures without making strong assumptions on the fault system geometry (e.g., Meng et al., 2012a) and that can be applied rapidly after large earthquakes as soon as teleseismic P-wave data are available (the results reported here were ready for our analysis on February 7th, see Ampuero (2023). The method can be automated to deliver results within 1 hour of any large earthquake). We used data from the Alaska array (Figure 2), which consists of 293 broadband seismic stations within 30 to 90° from the epicenter. Its high station density and large aperture allows an excellent spatial resolution in the source region. We employed the Multitaper-MUSIC back-projection method (Meng et al., 2011; Schmidt, 1986), which tracks the coherent sources of high frequency radiation with finer spatial resolution than conventional back-projection techniques. Our ray tracing for back-projection adopts the spherically symmetric IASP91 velocity model (Kennett and Engdahl, 1991). We first aligned the vertical components of the initial P waves using a standard iterative, cross-correlation technique (Reif et al., 2002) to correct for the travel time error caused by horizontal variations of velocity structure. We selected only the waveforms with a correlation coefficient greater than 0.6 to increase the waveform coherence and improve result quality. We then filtered the seismograms from 0.5 to 2 Hz and applied back-projection on a sliding window of 10 s. The back-projection is relative to the earthquake epicenter reported by the USGS, which is located off the EAF, in agreement with that determined by the AFAD (Figure 1).

The back-projection results provide a first-order view of the multi-fault rupture pattern. Figure 2a shows the resulting locations of high-frequency radiators. They coincide well with the active faults. The back-projection results reveal that the rupture initially propagated to the NE for the first 20 s, along a strike direction consistent with the splay fault. After reaching the junction with the EAF, the rupture became bilateral, propagating NE-ward and SW-ward simultaneously along the EAF, until the earthquake terminated at approximately 70 s. Due to the limited resolution and possibly interference between waves from multiple rupture fronts, it is challenging to determine whether the SW-ward rupture initiated simultaneously with the NE-ward rupture or after a delay. Although this issue may be resolved by later observational studies, the current ambiguity motivates us to examine a range of possible scenarios in Section 3. In any case, the back-projection results clearly confirm a pattern of rupture branching from the splay fault to the EAF, with a forward component to the NE and a backward component to the SW.

The back-projection results also provide constraints on rupture speed. Figure 2b illustrates that the rupture speed is approximately 3.2 to 3.5 km/s, indicating an overall subshear rupture (the shear wave speed $V_S$ at the depth of 8 to 10 km is ~ 3.15 to 3.6 km/s, Delph et al. (2015)). However, Rosakis et al. (2023) proposed an early supershear transition on the splay fault ~ 20 km away from the epicenter, based on the relative amplitudes of the fault-parallel and fault-normal components of near-field seismic recordings. Their estimated instantaneous rupture speed is approximately $1.55 \times V_S$ (or ~ 4.88 to 5.58 km/s). Unfortunately, the spatial and temporal scales of the proposed supershear rupture are smaller than the resolutions of teleseismic back-projection. Nonetheless, the average rupture speed of 3.2 to 3.5 km/s resolved by back-projection suggests that the proposed supershear rupture on the splay fault probably did not persist for long, if it indeed occurred, likely due to the impeding effect of the intersection with the EAF. A later finite source inversion study achieved detailed modeling of the recordings near the splay fault with a subshear rupture (Delouis et al., 2023). To cover different possible situations, we explore both subshear and supershear rupture speeds in our numerical simulations in Section 3.

## 2.3. Strong ground motion observations

The strong ground motion data also provide a first-order constraint on the rupture process, especially during the early stage. We retrieved the acceleration waveforms for a selected set of stations (Figure 3a) from the AFAD, with additional corrections for instrument response and baseline. We then filtered the corrected waveforms in the 1-20 Hz frequency band, and subsequently obtained the ground motion amplitude as $A = \sqrt{N^2 + E^2 + Z^2}$, where $N$, $E$, and $Z$ are the north, east, and vertical components of the recordings. As shown in Figure 3a, the nearest stations to the southwest (4615), west (4624, 4625, and 4616), north (4611 and 4631) and northeast (0208 and 0213) of the junction between the splay fault and the EAF permit a rough estimation of the early rupture process around the fault junction. By aligning these stations in the east-west direction relative to the fault junction, two groups of signals can be observed for those stations located in the west (Figure 3b). First, the onsets of large ground motion amplitudes recorded prior to 20 s since the origin time coincide with the P- and S-wave arrivals emanating from the hypocenter. Second, an additional phase of strong motions is observed >20 s after the origin time, which we interpret to originate from the passage of the rupture front. Assuming that the initial rupture arrived at the junction at around 16 s, we argue that a new rupture initiated along the EAF and then propagated bilaterally to the NE and SW, according to the moveouts of coherent high-frequency signals later than 20 s (indicated by the dashed green lines in Figure 3b). A simple estimation, based on the onsets of these late signals, yields a propagation speed of ~ 3.5 km/s along the EAF west of the junction. Although uncertainties still remain about the exact initiation location and earlier propagation speed of the rupture along the EAF, the strong ground motion data confirm a pattern of bilateral rupture branching from the splay fault to the EAF, consistent with the back-projection results.

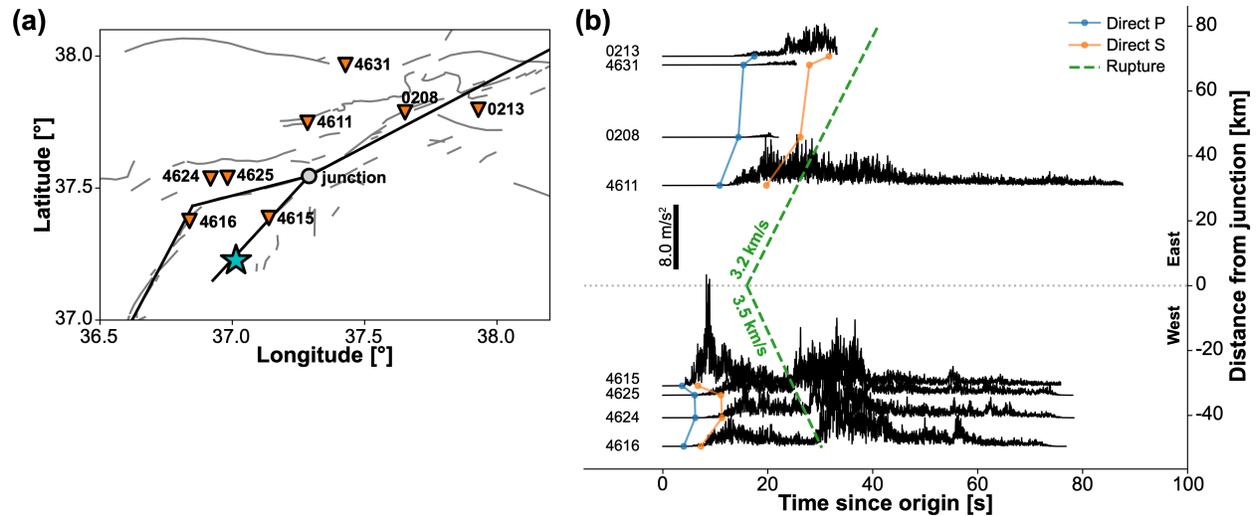

*Figure 3. Distribution map of strong ground motion stations and estimation of rupture process. (a) Discretization of the fault segments for the strong ground motion analysis and location of the stations included in the analysis. (b) Strong ground motion amplitudes (with a 1-20 Hz bandpass filtering) recorded by the stations around the fault junction. The distances are measured as the projection onto the nearest fault shown in (a) and are relative to the junction. The predicted direct P and S-arrivals emanating from the hypocenter and the average rupture trajectory are indicated by the solid and dashed lines, respectively.*

## 3. Dynamic rupture models explaining the observed rupture path

To better understand the observed rupture branching pattern, we conduct numerical simulations of dynamic ruptures exploring a range of possible scenarios, taking into account the uncertainty and diversity of rupture properties (e.g., fault slip, first triggered location(s) on the EAF, rupture speed) reported by different studies (Delouis et al., 2023; Melgar et al., 2023; Okuwaki et al., 2023; Rosakis et al., 2023).

### 3.1. Model settings

We build the fault model based on the surface rupture trace and aftershock distribution of the $M_w$ 7.8 mainshock, focusing on the area of the junction of the splay fault and the EAF (Figure 1). This region roughly falls into the Pazarcık segment of the EAF. Hereafter, we refer to the EAF in the numerical model as the main fault. Specifically, we consider a simplified model, where both the splay fault and the main fault are assumed to be planar and dipping vertically. For convenience, we rotate the view to define a Cartesian coordinate system aligned with the main fault (Figure 4). We set the angle $\theta$ between the two faults at 30°, based on the aftershock distribution (Figure 1). According to the results of stress inversion from historic seismicity (Güvercin et al., 2022), the regional stress orientation changes systematically along the strike of the EAF. For the Pazarcık segment, we set the angle $\Psi$ between the maximum compressive stress $\sigma_{max}^0$ and the main fault at 40°. Assuming fluid overpressure (Rice, 1993) and after some trial tests, we consider a regionally uniform initial stress field: $\sigma_{xx}^0 = -57.05$ MPa, $\sigma_{xy}^0 = -20$ MPa, $\sigma_{yy}^0 = -50$ MPa (negative sign for compression or left-lateral shear). Since our focus is on how rupture can propagate from the splay fault to the main fault, we truncate the main fault at a total length of 220 km, roughly centered at the junction, ignoring subsequent rupture propagation further to the NE and SW. We initiate the rupture along the splay fault, setting the hypocenter (red star in Figure 4) at 35 km from the fault junction. In the southward portion of the splay fault (dashed grey line in Figure 4), we assume a fault cohesion of 10 MPa to artificially terminate the rupture at around 20 km from the hypocenter.

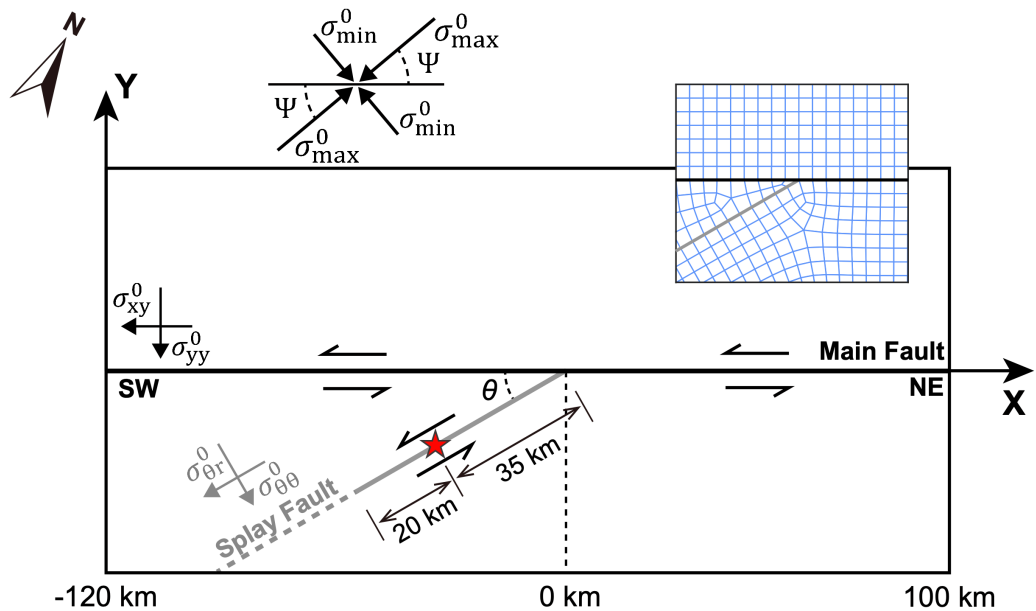

*Figure 4. Model setup for numerical simulations. The main fault (horizontal black line) mimics the EAF, while the splay fault (inclined grey line) mimics the short fault segment that hosted the hypocenter (red star). The angle $\theta$ between the two faults is set at 30°. A time-weakening friction is used to nucleate the rupture inside a finite-length zone along the splay fault. Elsewhere, a linear slip-weakening friction is*

*adopted. A fault cohesion of 10 MPa is set to terminate the southward rupture along the splay fault (dashed grey line). The angle Ψ between the maximum compressive stress $\sigma_{max}^0$ and the main fault is set at 40°. Off-fault materials are assumed linearly elastic. Absorbing boundary conditions are applied to the four edges of the domain. The inset shows the spectral element mesh near the fault junction. Other specifications of the numerical model can be found in the main text.*

We discretize the simulation domain with a quadrilateral element mesh generated in a previous study (Xu et al., 2015) with the software CUBIT (https://cubit.sandia.gov/). The elements have a size of 200 m on average, which translates into a spatial resolution of about 50 m (there are multiple internal nodes within each element). The treatment of the fault junction follows the convention commonly adopted for fault branching problems (DeDontney et al., 2012; Xu et al., 2015), where split nodes run continuously across the junction along the main fault but converge to a single non-split node at the junction along the splay fault. Such treatment allows for through-going rupture along the main fault but terminated rupture along the splay fault (zero splay-fault slip at the junction), which is supported by the relative maturity of the main fault (the EAF) and source inversion results (Melgar et al., 2023; Okuwaki et al., 2023).

We use a time-weakening friction with prescribed rupture speed of 2 km/s to artificially nucleate the rupture along the splay fault (Andrews, 1985; Bizzarri, 2010). Once the rupture exceeds a critical length, it spontaneously transitions to a linear slip-weakening friction law (Palmer and Rice, 1973; Andrews, 1976) in which the friction coefficient $f$ depends on slip $\Delta u$ as

$$f = \begin{cases} f_s - (f_s - f_d)\,\Delta u/D_c\,, & \text{if } 0 \le \Delta u \le D_c \\ f_d, & \text{if } \Delta u > D_c \end{cases} \tag{1}$$

where $f_s$, $f_d$, and $D_c$ are the static friction coefficient, dynamic friction coefficient and critical slip-weakening distance, respectively. Due to the lack of near-field dynamic stress measurements, we cannot directly constrain $f_s$ and $f_d$; we thus set their values based on trial and error. In contrast, there are many strong ground motion stations near the ruptured faults (Figures S1-S3), with which we estimate $D_c$ following the approach of Fukuyama and Mikumo (2007). In that approach, $D_c$ is estimated by a proxy $D_c'$ defined as two times the fault-parallel displacement at the time of peak ground velocity measured directly at the fault surface. In practice, stations are at some distance from the fault and we measure $D_c''$, an off-fault estimate of $D_c'$ (Figure S2). Numerical experiments indicate that $D_c''$ increases (Cruz-Atienza et al., 2009) with distance from the fault and thus provides an upper bound on $D_c'$. The resulting values of $D_c''$, shown along strike in Figure S3, range from 1.0 to 2.0 m. We note a heterogeneous distribution of $D_c''$ in the southern portion of the EAF, with lower values at 90-110 km from the junction. In numerical simulations, we explore $D_c$ values in the range of 0.5-2.0 m, as a compromise between observational constraint and computational cost. Smaller $D_c$ values are possible but require finer numerical resolution and hence increased computational cost. Larger $D_c$ values in general do not favor successful rupture branching and hence can be readily ruled out.

Based on the assumed initial stresses and friction coefficients, we calculate the seismic $S$ ratio to judge the relative closeness to failure and hence the rupture mode along each fault (Andrews, 1976; Das and Aki, 1977; Liu et al., 2014):

$$S = \frac{|\sigma_0|f_s - |\tau_0|}{|\tau_0| - |\sigma_0|f_d} \tag{2}$$

where $\tau_0$ and $\sigma_0$ are the initial shear and normal stress resolved on the faults. The parameter $S$ generally has a strong control on the properties of single-fault ruptures, such as their rupture speed (Andrews, 1976). However, in our multi-fault rupture case, $S$ as defined in Eq. (2) can only roughly characterize the rupture mode along the main fault after branching, because the effective initial stress field for the main fault can be modified by the rupture along the splay fault (Xu et al., 2015).

Viscous damping (Day et al., 2005) and normal stress regularization (Rubin and Ampuero, 2007; Xu et al., 2015) are applied to both faults to stabilize the simulation. For simplicity, we assume the surrounding medium is elastic and hence ignore any permanent deformation off the faults. To take into account the finite width of the seismogenic zone while keeping the computational efficiency of 2-D modeling, we adopt the 2.5-D approximation as in Weng and Ampuero (2019, 2020). A parameter $W$ is introduced to mimic the fault width, which causes a saturation of slip once the along-strike propagation distance exceeds a value proportional to $W$ (Day, 1982). In this study, we choose $W = 40$ km, equivalent to a fault width of 20 km in a half space (Luo et al., 2017), to match the observations of co-seismic slip and aftershocks for the $M_w$ 7.8 mainshock (Barbot et al., 2023; Melgar et al., 2023; Okuwaki et al., 2023). We carefully choose the values for other model parameters so that the simulated slip is comparable to the one found in source inversions. Unless mentioned otherwise, we assume model parameters are uniformly distributed along each fault. Their specific values may vary from one simulation case to another. Table 1 summarizes the key model parameters and their values. Superscripts $sp$ and $m$ indicate parameters along the splay fault and the main fault, respectively. We conduct the dynamic rupture simulations with the spectral-element-method software SEM2DPACK (Ampuero, 2012).

Table 1. Model parameters and their values

| Parameters | Values |
|---|---|
| Shear modulus $\mu$ | 32.4 GPa |
| P-wave speed $V_P$ | 6000 m/s |
| S-wave speed $V_S$ | 3464 m/s |
| Rayleigh-wave speed $V_R$ | 3185 m/s |
| Critical rupture speed $V_{rc}$ to define the end of nucleation | 1000 m/s |
| Width of the seismogenic zone (full space) $W$ | 40 km |
| Angle between maximum compressive stress and the main fault $\Psi$ | 40° |
| Angle between the main fault and splay fault $\theta$ | 30° |
| Initial normal stress along $x$ direction $\sigma_{xx}^0$ | −57.05 MPa |
| Initial normal stress along $y$ direction $\sigma_{yy}^0$ | −50 MPa |
| Initial shear stress $\sigma_{xy}^0$ | −20 MPa |
| Initial normal stress along the splay fault $\sigma_{\theta\theta}^0$ | −34.4 MPa |
| Initial shear stress along the splay fault $\sigma_{\theta r}^0$ | −6.95 MPa |
| Static friction coefficient along the splay fault $f_s^{sp}$ | 0.21~0.35 |
| Dynamic friction coefficient along the splay fault $f_d^{sp}$ | 0.10 |
| Critical slip-weakening distance along the splay fault $D_c^{sp}$ | 0.50~1.00 m |
| Static friction coefficient along the main fault $f_s^m$ | 0.42~0.48 |
| Dynamic friction coefficient along the main fault $f_d^m$ | 0.10~0.35 |
| Critical slip-weakening distance along the main fault $D_c^m$ | 0.50~2.00 m |

## 3.2. Delayed rupture triggering on the SW segment of the main fault

We first show the spatiotemporal evolution of Coulomb failure stress changes ΔCFS for a case of successful rupture branching from the splay fault to the main fault (Figure 5). Here, ΔCFS is defined as:

$$\Delta\text{CFS} = \Delta\tau + f^{eff} \times \Delta\sigma_n \qquad (3)$$

where $\Delta\tau$ and $\Delta\sigma_n$ are respectively the shear and normal stress changes induced by the splay-fault rupture, and $f^{eff}$ is an effective friction coefficient that includes the contribution from pore fluid pressure (Freed, 2005). In the plots, $\Delta\text{CFS}$ is either projected along the main fault (Figure 5a), or resolved onto planes parallel to the main fault (Figure 5b-e). Positive $\Delta\text{CFS}$ indicates increased chance for triggering left-lateral slip along faults parallel to the main fault.

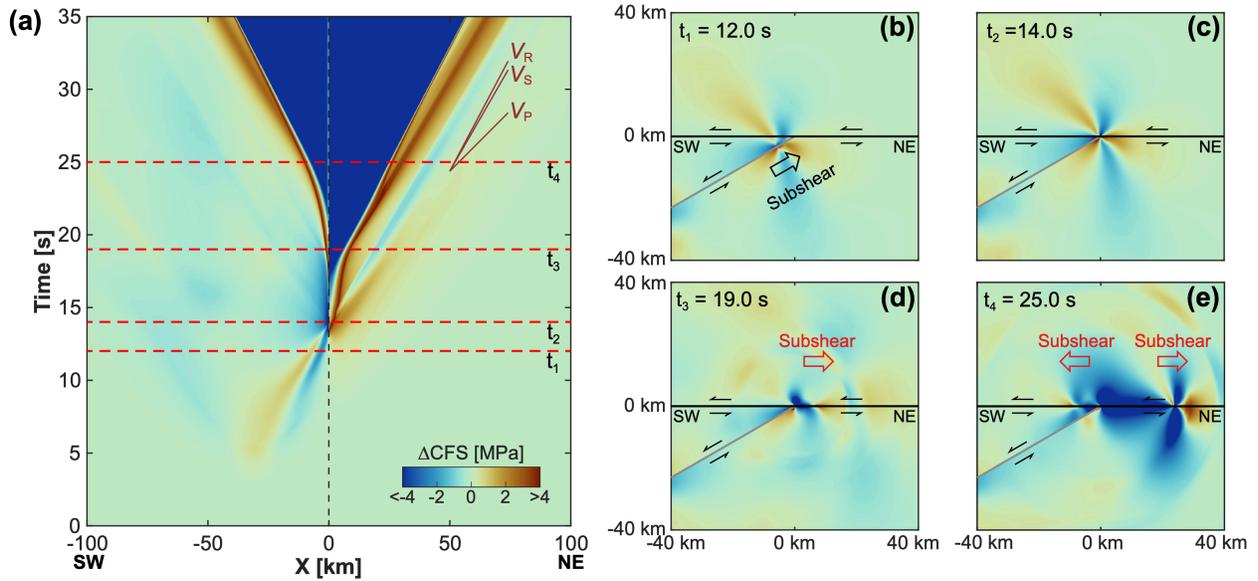

Figure 5. Spatiotemporal distribution of Coulomb failure stress change ($\Delta CFS$, positive promoting left-lateral shear) induced by a subshear rupture along the splay fault. (a) Evolution of $\Delta CFS$ along the main fault. Four times $t_1$ to $t_4$ are selected to highlight: (1) when a transient positive $\Delta CFS$ lobe can operate on the SW segment of the main fault, (2) when rupture just hits the junction (X = 0 km) along the splay fault, (3) when the NE segment of the main fault starts to slip, and (4) when the SW segment of the main fault starts to slip. (b)-(e) Spatial distribution of $\Delta CFS$, resolved onto faults parallel to the main fault, at the four selected times defined in (a). $f^{eff} = 0.48$ is assumed for computing $\Delta CFS$ (Eq. 3). Other model parameters are: $f_s^{sp} = 0.28$, $f_d^{sp} = 0.10$, $D_c^{sp} = 1.00\ m$; $f_s^m = 0.48$, $f_d^m = 0.29$, $D_c^m = 1.00\ m$.

In Figure 5, rupture initially propagates at subshear speed along the splay fault. As the rupture front approaches the fault junction, a positive $\Delta$CFS lobe sweeps over the SW segment of the main fault, moving towards the fault junction (before and around $t_1$ in Figures 5a and 5b). However, the amplitude of this positive $\Delta$CFS lobe is not strong enough to trigger slip on the main fault. Once the splay-fault rupture arrives at the junction, positive and negative $\Delta$CFS lobes persistently operate on the NE and SW segments of the main fault, respectively ($t_2$ to $t_3$ in Figures 5a, c and d). These two stress lobes with opposite signs are caused by the terminated rupture along the splay fault. They are long-lived and hence modify the effective initial stress of the main fault, promoting left-lateral slip on the NE segment and suppressing left-lateral slip on the SW segment. Indeed, a new rupture along the main fault is triggered around $t_3$ on the NE side of the junction, and starts to propagate to the NE (Figure 5d). Only after this rupture propagates beyond some distance, it transfers enough stress to the opposite side of the junction (Tada et al., 2000), allowing the SW segment to overcome the initial stress shadow ($t_2$ to $t_3$ in Figure 5a) and to finally start propagating in the SW direction ($t_4$ in Figures 5a and 5e).

Therefore, for the case in Figure 5, backward rupture branching is not achieved by a direct rupture branching from the splay fault to the SW segment of the main fault (Fliss et al., 2005), but through a cascade process from the splay fault to the NE segment of the main fault and then to the SW segment of the main fault. If the NE segment is forced to remain locked, then the SW segment is not successfully triggered (not shown here but confirmed by simulations), at least when the seismic S ratio on the main fault is not extremely low. We also test another case (Figure S4) with supershear rupture along the splay fault, as proposed by Rosakis et al. (2023), and find the above conclusion still holds.

After understanding the basic process of rupture branching, we proceed to investigate other aspects of the simulated results. Figure 6 shows the evolutions of slip rate and slip for the case in Figure 5. Rupture is not instantaneously triggered along the main fault, but displays a slow nucleation phase (Ohnaka, 1992) before accelerating to a speed close to the Rayleigh wave speed $V_R$ (Figure 6a). The NE segment of the main fault is activated earlier and hosts a faster rupture than the SW segment (Figure 6a). This is consistent with our previous judgement that the NE-ward rupture serves as a prerequisite for the SW-ward rupture along the main fault.

The asymmetry in rupture behaviors along the main fault is also manifested in the slip distribution along the main fault (Figure 6b). First, slip starts to accumulate around 15 s, first on the NE side of the junction.

Second, the average slip is always larger along the NE segment than on the SW segment, despite the same initial stress and frictional properties along both segments. The asymmetry in slip distribution supports our earlier statement that the effective initial stress field for the main fault is modified by the rupture along the splay fault, featuring long-lived positive ΔCFS for the NE segment but negative ΔCFS for the SW segment ($t_2$ to $t_3$ in Figure 5a). Such asymmetric slip distribution is also observed in the case with supershear rupture along the splay fault (Figure S5), in the source inversion results for the $M_w$ 7.8 mainshock (Barbot et al., 2023; Melgar et al., 2023; Okuwaki et al., 2023), and in other studies of the rupture branching problem (Bhat et al., 2007a; Fliss et al., 2005; Templeton et al., 2009; Xu et al., 2015), suggesting that it should be a common feature around fault junctions (Andrews, 1989).

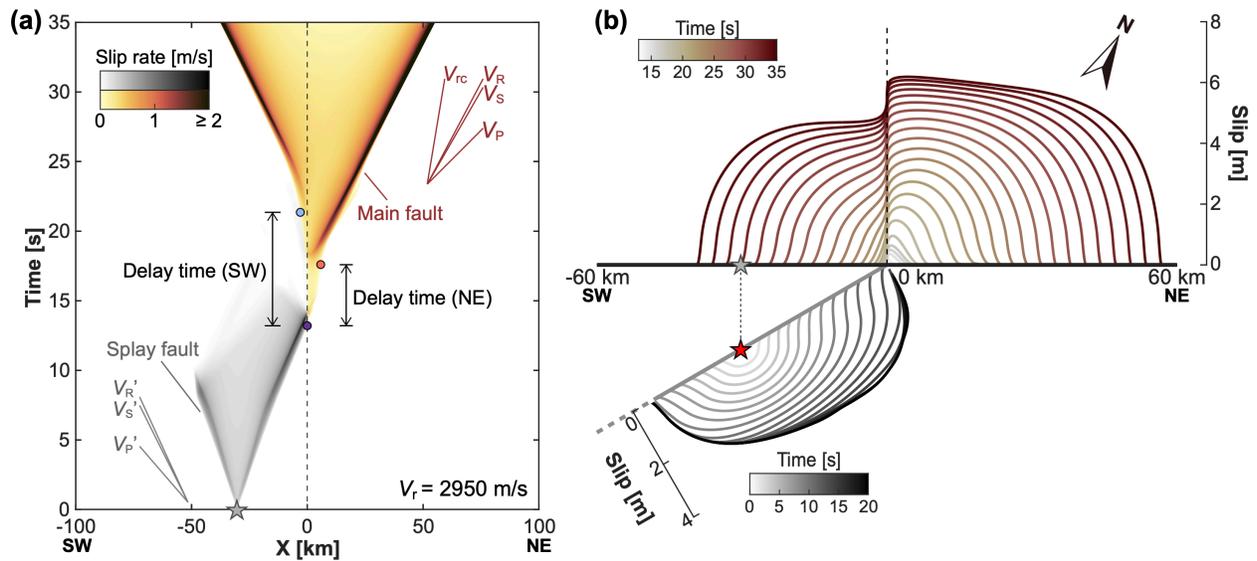

*Figure 6. Spatiotemporal distribution of (a) slip rate and (b) slip for the case shown in Figure 5. In (a), the evolution of slip rate along the splay fault is projected onto the plane parallel to the main fault. $V_P'$, $V_S'$ and $V_R'$ represent the apparent P-, S- and Rayleigh-wave speed projected onto the same plane. For the NE and SW segments of the main fault, the delay time is defined as the interval between when splay-fault rupture just arrives at the junction and when triggered main-fault rupture attains a critical propagation speed $V_{rc}$ of $1\ km/s$. $V_r$ ($2950\ m/s$) denotes the instantaneous speed of the splay-fault rupture prior to the arrival at the junction. In (a) and (b), grey star represents the projection of the rupture hypocenter (red star) onto the main fault.*

### 3.3. Controls on rupture triggering on the main fault

To provide quantitative understanding of what controls the rupture branching from the splay fault to the main fault, we have conducted two sets of numerical simulations. In the first set, we fix the parameters along the splay fault (same as in Figures 5 and 6), but vary those along the main fault, in particular the critical slip-weakening distance $D_c^m$ and the seismic S ratio $S^m$ (by varying the dynamic friction coefficient $f_d^m$). We investigate how main fault properties affect the triggering process, especially the delay time defined as the interval between when rupture arrives at the junction along the splay fault and when the rupture triggered along the main fault reaches a propagation speed $V_{rc}$ of 1 km/s (see the definition in Figure 6a). The value of $V_{rc}$ is arbitrary; nonetheless, the chosen value of 1 km/s does provide a reference for judging the end of the nucleation process along the main fault. According to the scenario in Figures 5 and 6, the main-fault rupture starts only after the splay-fault rupture arrives at the junction; therefore, the delay time is positive for both segments of the main fault.

Figure 7 summarizes the results of delay time for varying parameters along the main fault. The delay time is shorter for the NE segment than for the SW segment (compare Figures 7a and 7b), again supporting our previous judgement of NE-ward rupture as a prerequisite for the SW-ward rupture along the main fault (Figure 5). On each segment, the delay time, a proxy of rupture nucleation time, increases with the critical slip-weakening distance $D_c^m$ and the seismic S ratio $S^m$. Given that rupture length scales with time during the nucleation process, the observed trend is consistent with the theory on rupture nucleation under quasi-static loading (Uenishi and Rice, 2003), despite that in this study rupture can be dynamically triggered along the main fault. With a decrease of $S^m$, there is a transition of rupture mode from subshear to supershear along the main fault, which is in general agreement with the results of previous studies (Andrews, 1976; Liu et al., 2014). Moreover, the transition boundary usually occurs at larger $S^m$ for the NE segment (Figure 7b) than for the SW segment (Figure 7a). This again can be explained by the modification of the effective initial stress on the main fault: under an overall positive (or negative) $\Delta$CFS for the NE (or SW) segment (Figure 5), the actual seismic S ratio can be smaller (or larger) than the nominal one used in Figure 7b (or Figure 7a).

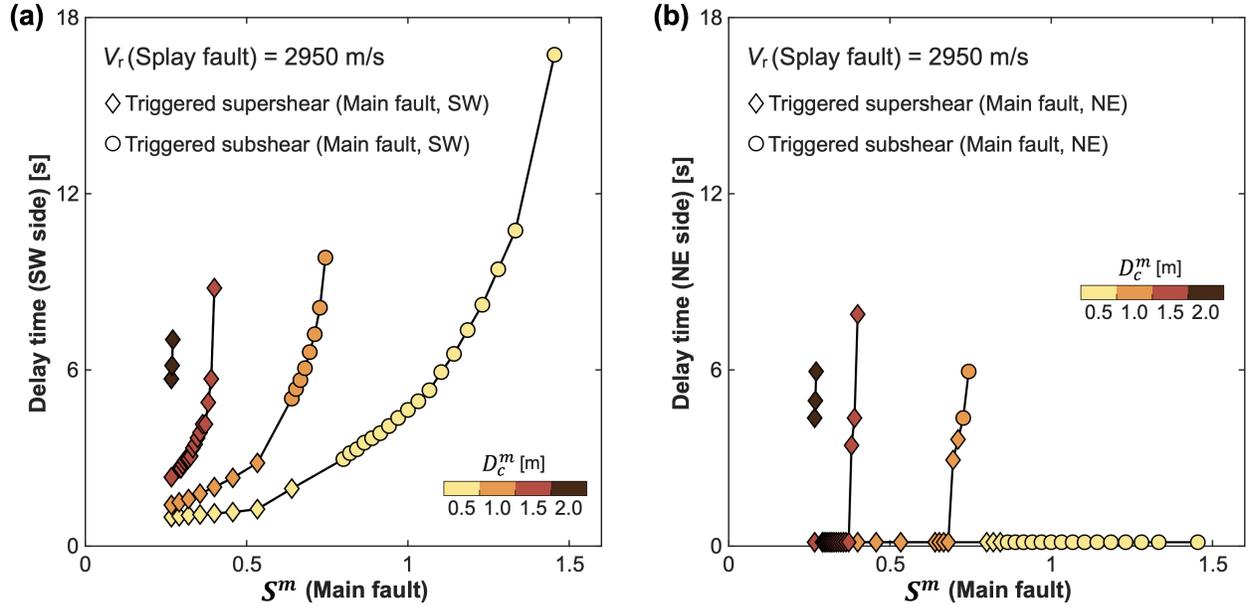

*Figure 7. Summary of delay time as a function of seismic S ratio ($S^m$) and critical slip-weakening distance ($D_c^m$) along the main fault. (a) and (b) are for the delay time along the SW and NE segment of the main fault, respectively. We fix the parameters along the splay fault ($f_s^{sp} = 0.28$, $f_d^{sp} = 0.10$, $D_c^{sp} = 1.00\ m$), resulting in a subshear rupture ($V_r = 2950\ m/s$, same as in Figure 6) prior to the arrival at the junction. For the main fault, we vary $f_d^m$ (under fixed $f_s^m = 0.48$) to obtain different values for $S^m$. Under this consideration, rupture can always be triggered along the main fault (at least for the NE segment), as predicted by the $\Delta CFS$ computation in Figure 5.*

In the second set of numerical simulations, we fix the parameters along the main fault (same as in Figures 5 and 6), but vary those along the splay fault. We first show three examples in Figure 8, from which three prominent features are observed. First, the triggered rupture along the main fault can be asymmetric, featuring supershear towards NE but subshear towards SW (Figure 8b and c), which can still be attributed to the asymmetric $\Delta CFS$ across the junction along the main fault. Second, with a decrease of rupture speed $V_r$ along the splay fault (from Figure 8a to 8c), the speed of the triggered rupture along the main fault increases, especially on the NE segment, which is unexpected. Third, the delay time is the shortest for the case in Figure 8c, despite having the slowest rupture speed along the splay fault. Our parametric study further confirms that the delay time tends to be shorter when rupture speed is slower (under larger $S^{sp}$) along the splay fault (Figure 9). To deepen the understanding of rupture triggering along the main fault, we further evaluate the loading applied by the splay fault rupture on the main fault. Specifically, we examine how shear stressing rate $|\dot{\tau}|$, evaluated at the splay-fault rupture front right before it hits the junction (Figure 10a-c), influences the rupture nucleation and propagation along the main fault.

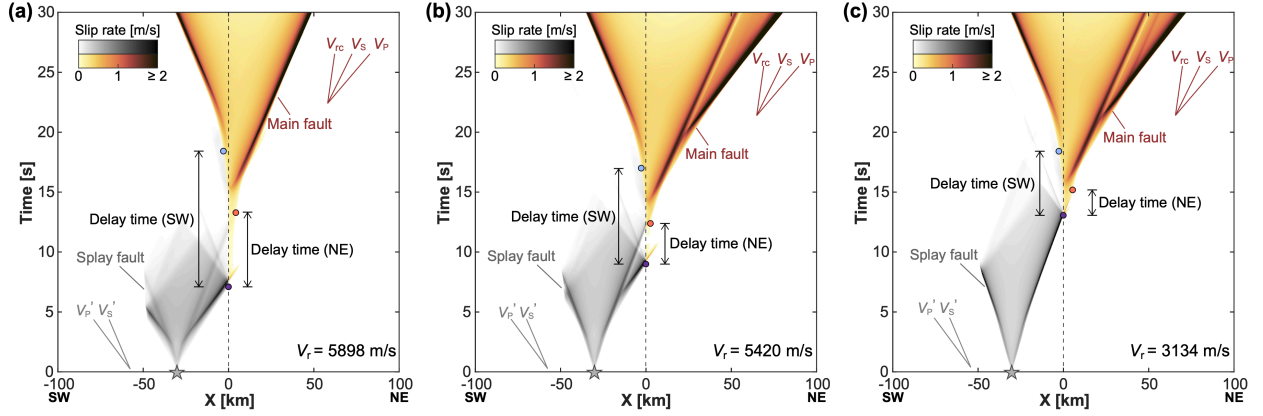

*Figure 8. Evolution of slip rate for three examples featuring diverse rupture behaviors. (a) Splay-fault rupture transitions to supershear at an earlier time; main-fault rupture remains subshear in both directions. (b) Splay-fault rupture transitions to supershear at a later time; along the main fault, the NE-ward rupture transitions to supershear while the SW-ward rupture remains subshear. (c) Splay-fault rupture remains subshear; along the main fault, the NE-ward rupture transitions to supershear while the SW-ward rupture remains subshear. For simulating the three examples, we fix the parameters along the main fault ($f_s^m = 0.48$, $f_d^m = 0.29$, $D_c^m = 1.00 \ m$) and some parameters along the splay fault ($f_d^{sp} = 0.10$, $D_c^{sp} = 0.50 \ m$). We vary the static friction coefficient along the splay fault to obtain different rupture behaviors: (a) $f_s^{sp} = 0.21$, (b) $f_s^{sp} = 0.25$, and (c) $f_s^{sp} = 0.33$. In (a)-(c). $V_r$ denotes the instantaneous propagation speed of splay-fault rupture prior to the arrival at the junction. We also pick the same moment, when splay-fault rupture is about to hit the junction, to evaluate the shear stressing rate $|\dot{\tau}|$ (to be shown in Figure 10).*

We find $|\dot{\tau}|$ partly explains the unexpected results of the rupture process along the main fault (Figures 8 and 9). First, $|\dot{\tau}|$ for a splay-fault subshear rupture is higher than that for a splay-fault supershear rupture (Figure 10d and e), at least for the cases investigated in this study, which can be attributed to the higher degree of stress singularity in the subshear regime (Freund, 1990). Higher $|\dot{\tau}|$ along the splay fault corresponds to higher loading rate along the main fault, based on the compatibility condition for elasticity.

Second, previous studies indicate that higher loading rate can reduce the time or length required for rupture nucleation and can promote faster rupture propagation beyond nucleation (Guérin-Marthe et al., 2019; Gvirtzman and Fineberg, 2021; Kato et al., 1992; McLaskey and Yamashita, 2017; Xu et al., 2018; Yu et al., 2002). Taken together, it becomes clear that subshear rupture along the splay fault can exert a higher loading rate to the junction region, which favors earlier triggering and faster rupture propagation along the main fault, at least for the NE segment where $\Delta$CFS remains positive (Figure 9b). Using a similar argument for the main fault, one can also explain the triggering of SW-ward rupture (Figure 9a) aided by the stress transfer from the NE segment. Loading rate is not the only factor that can affect the rupture process under dynamic loading. Loading duration (Xu et al., 2015), successive loading (e.g., due to supershear front, S-wave Mach front, and Rayleigh front) (Aben et al., 2016; Smith and Griffith, 2022; Xu and Ben-Zion, 2017) and arrest waves (Rubin and Ampuero, 2007; Ryan and Oglesby, 2014) can also play a role, but are beyond the scope of this study.

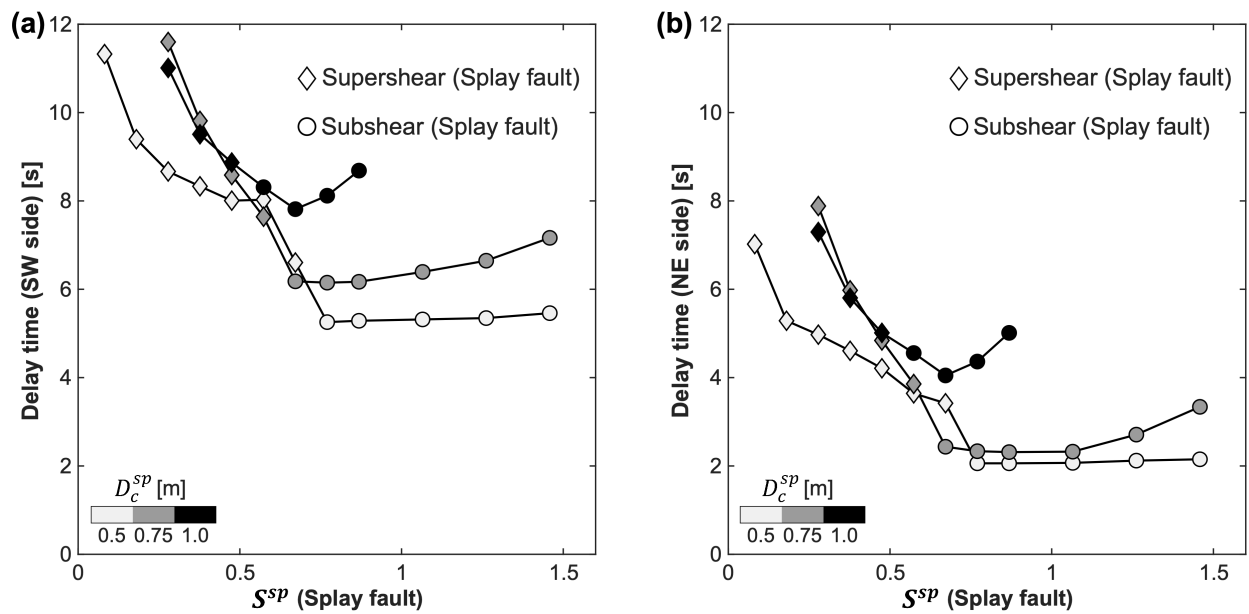

*Figure 9. Delay time as a function of seismic S ratio ($S^{sp}$) and critical slip-weakening distance ($D_c^{sp}$) along the splay fault. (a) and (b) are for the delay time along the SW and NE segment of the main fault, respectively. We fix the parameters along the main fault ($f_s^m = 0.48$, $f_d^m = 0.29$, $D_c^m = 1.00\ m$), which are the same as in Figures 5 and 6. For the splay fault, we vary $f_s^{sp}$ (under fixed $f_d^{sp} = 0.10$) to obtain different values for $S^{sp}$. Under this consideration, the simulated maximum slip ($\sim$3 m) along the splay fault remains comparable to the one inferred by source inversion of the $M_w$ 7.8 mainshock.*

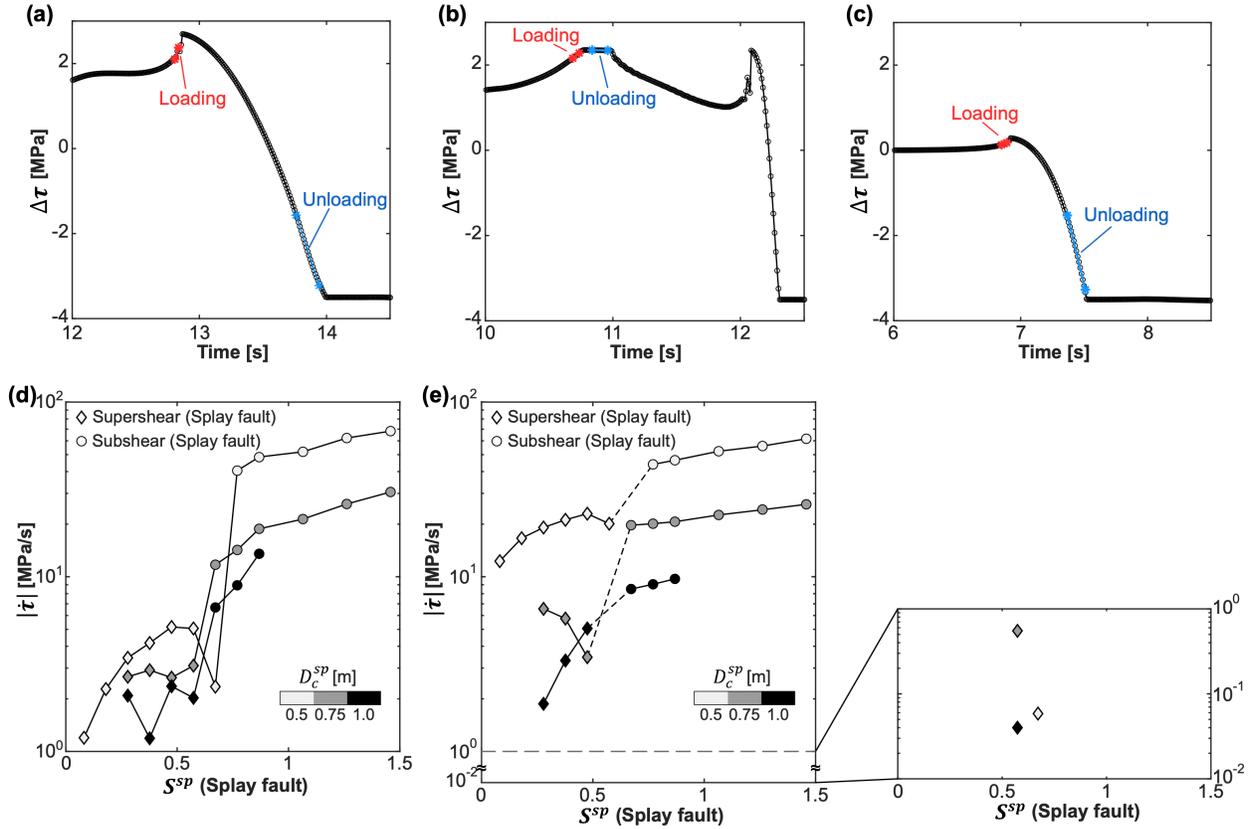

*Figure 10. Simulated shear stress waveforms for (a) a subshear rupture, (b) a supershear daughter rupture that is just born ahead of a subshear mother rupture, and (c) a supershear rupture that is already well established. In each panel of (a)-(c), two versions of shear stressing rate $|\dot{\tau}|$ are estimated, based on the loading (red) and unloading (blue) stage around the primary rupture front. More details about the three rupture modes in (a)-(c) can be found in Liu et al. (2014) and Xu et al. (2023). Summary of $|\dot{\tau}|$ during (d) loading and (e) unloading for a range of seismic S ratio ($S^{sp}$) and critical slip-weakening distance ($D_c^{sp}$) along the splay fault (same as in Figure 9).*

## 3.4. Early rupture triggering on the SW segment of the main fault

We also find cases in which rupture is triggered first on the SW segment of the main fault. Although we do not know whether this scenario indeed occurred during the mainshock, it is interesting to explore its features for the following three reasons. First, the splay-fault rupture exerts a transient positive ΔCFS on the SW segment of the main fault before the splay-fault rupture arrives at the junction (Figures 5 and S4). We aim at determining whether the amplitude and duration of this transient stress are enough for a successful triggering of the SW segment (Figures S6 and S7). Second, the last two large earthquakes (with a magnitude around or above 7) along the Pazarcık segment of the EAF occurred in 1795 ($M$ 7.0) and 1513 ($M_s$ 7.4) (Ambraseys, 1989; Güvercin et al., 2022). Therefore, it is possible that the Pazarcık segment (corresponding to our modeled main fault) was already close to failure before the 2023 $M_w$ 7.8 mainshock, which could permit an earlier triggering along its SW segment. Third, two prominent seismic clusters, associated with relatively low Gutenberg-Richter $b$-values, have been observed around the fault junction before the 2023 $M_w$ 7.8 mainshock (Kwiatek et al., 2023), suggesting that this region could have been already stressed close to failure before the mainshock.

Figure 11 shows one case in which the SW segment of the main fault is successfully triggered, before a subshear rupture arrives at the junction along the splay fault. Additional simulations (not shown here) confirm that the SW-ward rupture along the main fault can continue propagating even without activation of the NE segment. We find similar results for a supershear rupture along the splay fault (Figure S8). Our detailed investigation reveals that, as long as the main fault is initially close to failure (extremely low $S^m$), successful earlier triggering occurs along the SW segment of the main fault, leading to ruptures that can easily reach supershear speeds (Figure S9), are often characterized by a negative delay time (Figures 11a and S9), and appear to propagate "faster" (if mis-counted from the junction) than the NE-ward rupture (Figure 11b). Nonetheless, the final slip along the main fault is still smaller on the SW side of the junction (Figure 11b), similar to the previous case without earlier triggering on this side (Figure 6b). This may be explained by the fact that the final slip distribution along the main fault is more sensitive to the static ΔCFS, which is established only after the splay-fault rupture is terminated at the junction.

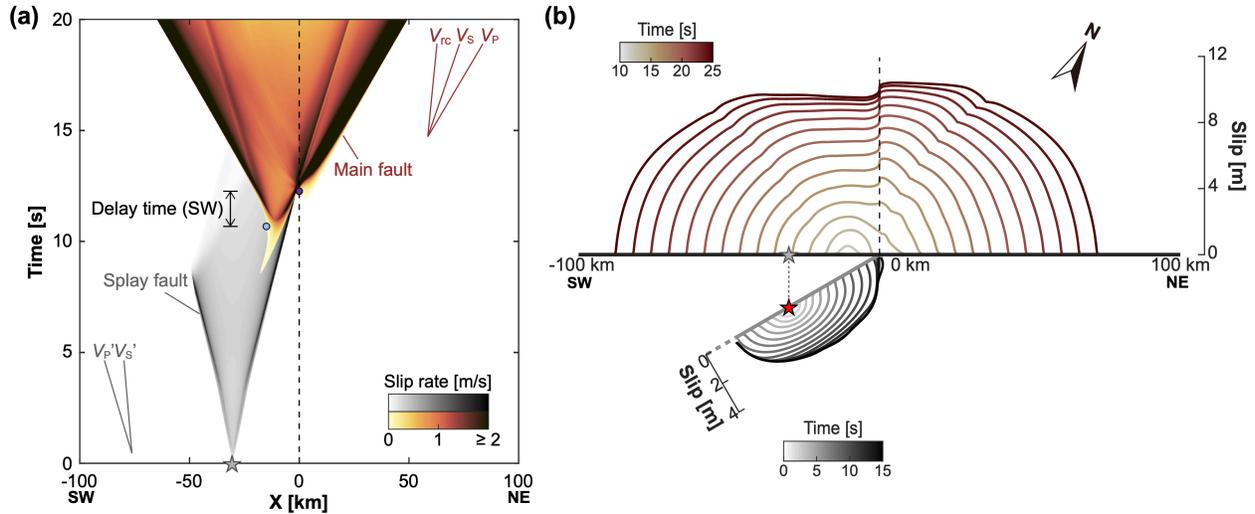

*Figure 11. Spatiotemporal distribution of (a) slip rate and (b) slip for a case with earlier triggering on the SW segment of the main fault. The splay-fault rupture remains subshear, while the main-fault rupture eventually reaches supershear in both directions. In (a), the delay time is still defined as the interval between the times when the splay-fault rupture just arrives at the junction and when the triggered main-fault rupture attains a propagation speed of $1\ km/s$. This delay time is now negative for the SW-ward rupture along the main fault. Due to the complex rupture behavior towards NE along the main fault, involving a second rupture triggering at the junction, we do not estimate the delay time for that rupture direction. The relevant model parameters are: $f_s^{sp} = 0.29$, $f_d^{sp} = 0.10$, $D_c^{sp} = 0.50\ m$; $f_s^m = 0.42$, $f_d^m = 0.24$, $D_c^m = 0.50\ m$.*

## 4. Discussion

### 4.1. The need to consider backward rupture branching in earthquake hazard assessment

Previous earthquake hazard analyses only considered scenarios of forward rupture branching to estimate the maximum magnitude of earthquakes in the Anatolian region (Mignan et al., 2015). By contrast, backward rupture branching occurred during the 2023 $M_w$ 7.8 Turkey earthquake, involving rupture propagation from a splay fault onto the SW segment of the EAF (Figures 1-3). According to our 2.5-D numerical simulations, such backward rupture branching can be realized in two different modes. In the first mode (Figures 5, 6, 8, S4 and S5), rupture does not make a direct transition from the splay fault to the backward segment of the main fault; rather, it triggers rupture along the forward segment, which in turn triggers rupture along the backward segment at a later time. In other words, a complex cascade process occurs sequentially over three fault segments, in which the intermediate segment plays a vital role in transferring positive $\Delta$CFS first in the forward direction and then in the backward direction. In the second mode (Figures 11, S6-S8), rupture on the splay fault directly triggers rupture along the backward segment of the main fault, if the latter is initially close to failure. Whether the forward segment of the main fault is present is not important, although its presence may facilitate the triggering of an additional rupture at a later time (Figures S7 and S8). In both modes, successful backward rupture branching can be realized for a range of rupture speeds along the splay fault (Figures 9 and S9).

Together with other known examples of backward rupture branching observed on strike-slip faults (Fliss et al., 2005; Li et al., 2020; Oglesby et al., 2003; see more examples in Xu, 2020), dip-slip faults (Xu et al., 2015, and references therein), as well as during laboratory earthquakes (Rousseau and Rosakis, 2003, 2009), our results suggest that backward rupture branching should be considered more systematically in earthquake hazard analyses. Future studies can be conducted to explore other conditions (e.g., 3-D effects, gap or overlap between different fault segments) that can promote or impede backward rupture branching. Efforts can also be made to classify the detailed situations for backward rupture branching, e.g., whether backward rupture branching is realized on pre-existing or newly-formed faults, with the same or opposite sense of slip, on the extensional or compressional side, and through a direct or indirect triggering process.

## 4.2. Anticipating rupture directivity

Rupture directivity exerts a first-order control on ground motion pattern, activation of secondary faults, and final earthquake size (Andrews and Ben-Zion, 1997; Lozos, 2016; Oglesby and Mai, 2012; Xu et al., 2015). Although modern observational networks, especially those installed near active faults, allow rupture directivity to be unambiguously determined, the number of instrumentally recorded earthquakes is still sparse. To gain confidence on the possible rupture paths in a given fault system, inferences are often drawn from a wealth of historic earthquakes, e.g., based on the fault geometry configuration (Fliss et al., 2005; Platt and Passchier, 2016; Scholz et al., 2010) and the permanent damage markers preserved in the field (Di Toro et al., 2005; Dor et al., 2006; Rowe et al., 2018). Taking the 2023 $M_w$ 7.8 Turkey earthquake as an example, based on the information of fault geometry (Figure 1) and the scenarios considered by Mignan et al. (2015), one would have expected the earthquake to nucleate in the middle of the Pazarcık segment of the EAF and then propagate bilaterally along different segments of the EAF, occasionally producing forward rupture branching in the extensional or compressional quadrants. However, the earthquake actually nucleated on a splay fault and then continued on the EAF, featuring a forward rupture branching to the NE and a backward rupture branching to the SW. Therefore, more works are needed to improve the methods to assess the possible rupture directivity of future events, especially when direct seismological constraints are not available.

**4.3. The impacts of splay-fault rupture on the main fault**

The initiation of the $M_w$ 7.8 Turkey earthquake on a splay fault raises several interesting questions about its possible impacts on the rupture of the main fault (the EAF). First, our numerical simulations show that, even with uniform initial stress and frictional properties, the rupture behavior on the main fault can be highly asymmetric across the junction, e.g., featuring a quicker rupture nucleation, a faster rupture speed and a larger slip on the NE side than on the SW side (Figures 6 and 8). According to our stress analysis (Figures 5 and S4), such asymmetric rupture behavior can be explained by the asymmetric stress change imposed by the splay-fault rupture. Without the latter, if the earthquake had started on the EAF, the evolution of rupture would have been smooth and continuous along the main fault, unless other complexities (e.g., nonlinear dynamics, spatial heterogeneity, inherent discreteness) are invoked (Cochard and Madariaga, 1996; Madariaga, 1979; Rice and Ben-Zion, 1996). Second, our numerical simulations also show that sometimes multiple ruptures can be almost simultaneously nucleated along the main fault (Figures S7 and S8), which is otherwise difficult to achieve under a slowly increased background loading unless strong heterogeneities are involved (Albertini et al., 2021; Cattania and Segall, 2021; Lebihain et al., 2021; McLaskey, 2019; Schär et al., 2021; Selvadurai et al., 2023; Yamashita et al., 2022). Again, such seemingly surprising result (no strong heterogeneities in our simulations) can be explained by the high-rate dynamic loading imposed by the splay-fault rupture, which is known to be capable of nucleating multiple ruptures (Doan and d'Hour, 2012), sometimes even with supershear speed (Xu et al., 2018, 2023). Finally, a third point can be raised by considering the failure time of the main fault and the associated earthquake size. Without the transient and static stress perturbations imposed by the splay-fault rupture, the main fault would have failed later (Gomberg et al., 1998), after accumulating additional strain energy, potentially leading to larger slip and a faster rupture speed. Alternatively, the high-rate dynamic loading imposed by the splay-fault rupture might have promoted the main-fault rupture to attain a very fast speed from the beginning (Guérin-Marthe et al., 2019; Gvirtzman and Fineberg, 2021; Kato et al., 1992;

McLaskey and Yamashita, 2017; Xu et al., 2023; Yu et al., 2002), allowing it to expand further than ever before (Güvercin et al., 2022) under a rate-dependent feedback mechanism (Xu et al., 2018), despite that the failure time has been advanced: rate-enhanced rock brittleness and co-seismic weakening could overtake the shortened healing time, leading to larger slip and a faster rupture speed (Hatakeyama et al., 2017; McLaskey, 2019).

In short, although the actual rupture behavior along the EAF during the $M_w$ 7.8 Turkey earthquake remains to be refined by ongoing observational studies, all three points above suggest that the rupture pattern on the EAF could have been different if the earthquake had started on the EAF as a result of slow tectonic loading. The same points may also apply to other regions (e.g., Baja and southern California) where a large earthquake is inferred to have started on a subsidiary fault (Fletcher et al., 2016; Lozos, 2016).

### 4.4. Back-propagating rupture mediated by fault geometry

If the observations of the 2023 $M_w$ 7.8 Turkey earthquake had been too coarse to resolve that it initiated on a splay fault, its backward rupture branching would have been interpreted as a case of back-propagating rupture, in which the rupture first propagated to the NE on the EAF and then turned around to the SW on the same fault. Back-propagating rupture has been reported in slow earthquakes (Houston et al., 2011; Obara et al., 2012), regular earthquakes (Hicks et al., 2020; Ide et al., 2011; Meng et al., 2012b) and laboratory earthquakes (Gvirtzman and Fineberg, 2021; Xu et al., 2023). Multiple mechanisms have been proposed to explain its occurrence: stress transfer along a heterogeneous fault (Luo and Ampuero, 2017), pore-pressure wave (Cruz-Atienza et al., 2018), low-velocity fault damage zone (Idini and Ampuero, 2020), free-surface reflection (Oglesby et al., 1998). Xu et al. (2021) argued that back-propagating rupture is an intrinsic feature of dynamic ruptures, whose observability is usually masked by the superposition effect but can be enhanced by various types of perturbation. For the previously reported cases, it was either observed or assumed that the rupture propagated back and forth along the same fault, and quite often the back-propagating rupture propagated faster than the forward rupture (Houston et al., 2011; Obara et al., 2012). However, this is clearly not the case for the Turkey earthquake, where at least two distinct faults were involved in the back-and-forth rupture propagation (Figures 1-3). Moreover, the back-propagating rupture could have been slower than the initial forward rupture, according to our (Figures 6 and 8) and other simulation results (Abdelmeguid et al., 2023). The multi-segment fault geometry plays the most important role in exciting the back-propagating rupture during the Turkey earthquake. Since multiple fault strands and triple junctions are common (Faulkner et al., 2003; Platt and Passchier, 2016; Rowe et al., 2013; Şengör et al., 2019; Vannucchi et al., 2012; Wolfson-Schwehr and Boettcher, 2019), some of the previously-reported back-propagating ruptures might have also been mediated by a multi-fault geometry that was not resolved in the available observations. Future efforts could aim at improving the methods for imaging fault zone structure and earthquake rupture processes, to assess the importance of fault geometry in back-propagating ruptures or to compare back-propagating ruptures on single faults and on multiple faults.

## 5. Conclusions

Motivated by the multi-segment rupture observed during the 2023 $M_w$ 7.8 Turkey earthquake, we have conducted 2.5-D numerical simulations of dynamic ruptures in a splay-and-main fault system, with rupture initiation on the splay fault. In particular, we focused on processes enabling the unexpected feature of backward branching involving rupture propagation from the splay fault to the southwest segment of the EAF, which makes an acute angle to the splay fault and lies in its static stress shadow. The simulated results show that bilateral rupture branching onto the main fault (representing the East Anatolian Fault, EAF) can be realized by both subshear and supershear ruptures along the splay fault. Two distinct modes of the branching process are identified. In the first mode, rupture branches from the splay onto the forward (NE) segment of the main fault which, after some delay, triggers the backward (SW) segment, revealing a complex cascade process across three fault segments. In the second mode, the backward segment of the main fault is directly activated by the splay-fault rupture, provided that the main fault is initially close to failure. While our numerical model is simplified in many aspects, including fault geometry, initial stresses and friction properties, the simulation results provide useful insights for understanding possible scenarios of rupture branching in configurations similar to the 2023 $M_w$ 7.8 Turkey earthquake. Especially, our study suggests that backward rupture branching, a feature deemed implausible by previous studies, should be considered in earthquake hazard analyses.

## Acknowledgments

SX was supported by the National Key R&D Program of China 2021YFC3000700 and the NSFC grant 42074048. YX and JPA were supported by the EU project "DT-GEO, A Digital Twin for Geophysical Extremes" (No 101058129). JPA was also supported by the French government through the UCA-JEDI Investments in the Future project (ANR-15-IDEX-01) managed by the National Research Agency (ANR). MvdE was supported by the European Research Council (ERC) under the European Union's Horizon 2020 research and innovation program (grant agreement No. 101041092 – ABYSS). JP was supported by fellowships from the Interdisciplinary Institute for Artificial Intelligence 3IA Côte d'Azur.

**Author contributions**

*Project design and supervision*: J. P. Ampuero and S. Xu. *Back-projection analysis*: Y. Xie. *Strong ground motion data analysis*: M. van den Ende and J. Premus. *Numerical simulations*: X. Ding. *Original draft writing*: All authors.

**Competing interests**

There are no competing interests.

**Data and code availability**

The software SEM2DPACK is freely available at: https://github.com/jpampuero/sem2dpack. The key input parameters for running numerical simulations are within the paper. The information for the $M_w$ 7.8 mainshock can be found from the AFAD (Turkey Disaster and Emergency Management Authority) at: https://deprem.afad.gov.tr/event-focal-mechanism, or from the USGS (U.S. Geological Survey) at: https://earthquake.usgs.gov/earthquakes/eventpage/us6000jllz/origin/detail. The aftershock catalog was downloaded from the AFAD at: https://deprem.afad.gov.tr/event-catalog (last accessed on May 29, 2023). The surface rupture trace was from the USGS (Reitman et al., 2023) at: https://doi.org/10.5066/P985I7U2. The teleseismic data of the Alaska array were downloaded through the IRIS Wilber 3 system (https://ds.iris.edu/wilber3/) including the following seismic networks: (1) the AK (Alaska Earthquake Center, Univ. of Alaska Fairbanks. (1987). Alaska Geophysical Network [Data set]. International Federation of Digital Seismograph Networks. https://doi.org/10.7914/SN/AK); (2) the AT (NOAA National Oceanic and Atmospheric Administration (USA). (1967). National Tsunami Warning Center Alaska Seismic Network [Data set]. International Federation of Digital Seismograph Networks. https://doi.org/10.7914/SN/AT); (3) the AV (Alaska Volcano Observatory/USGS. (1988). Alaska Volcano Observatory [Data set]. International Federation of Digital Seismograph Networks. https://doi.org/10.7914/SN/AV); (4) the CN (Natural Resources Canada (NRCAN Canada). (1975). Canadian National Seismograph Network [Data set]. International Federation of Digital Seismograph Networks. https://doi.org/10.7914/SN/CN); (5) the II (Scripps Institution of Oceanography. (1986). Global Seismograph Network-IRIS/IDA [Data set]. International Federation of Digital Seismograph Networks. https://doi.org/10.7914/SN/II); (6) the IM (Various Institutions. (1965). International Miscellaneous Stations [Data set]. International Federation of Digital Seismograph Networks. https://doi.org/10.7914/vefq-vh75); (7) the IU (Albuquerque Seismological Laboratory/USGS. (2014). Global Seismograph Network (GSN-IRIS/USGS) [Data set]. International Federation of Digital Seismograph Networks. https://doi.org/10.7914/SN/IU); (8) the US (Albuquerque Seismological Laboratory (ASL)/USGS. (1990). United States National Seismic Network [Data set]. International Federation of Digital Seismograph Networks. https://doi.org/10.7914/SN/US). The strong ground motion data were retrieved from the AFAD (https://doi.org/10.7914/SN/TK), using https://tadas.afad.gov.tr/login and https://tadas.afad.gov.tr/list-event.

**Supplementary material**

**S1. Estimation of $D_c$ based on $D_c{''}$**

Figures S1-S3 show the information for estimating the critical slip-weakening distance $D_c$ and the obtained results. We use unfiltered seismograms with baseline correction: we remove the mean value of acceleration recordings before the earthquake and a fitted quadratic function from velocity. As mentioned in the main text, $D_c$ is not directly measured but estimated by a proxy $D_c{''}$, defined as two times the fault-parallel displacement at the time of peak ground velocity measured at short distance from the fault (Figure S2). The proxy $D_c{''}$ is an upper bound of the value $D_c{'}$ that would be measured exactly on the fault, and $D_c{'}$ itself is a representative approximation of the actual $D_c$. We use the data retrieved from the entire portion of the EAF southwest of the junction with the initial splay fault (Figures S1 and S3), whereas in our numerical simulations we only focus on the region near the junction. Figure S3 shows our $D_c''$ estimates, their uncertainties and their spatial variability.

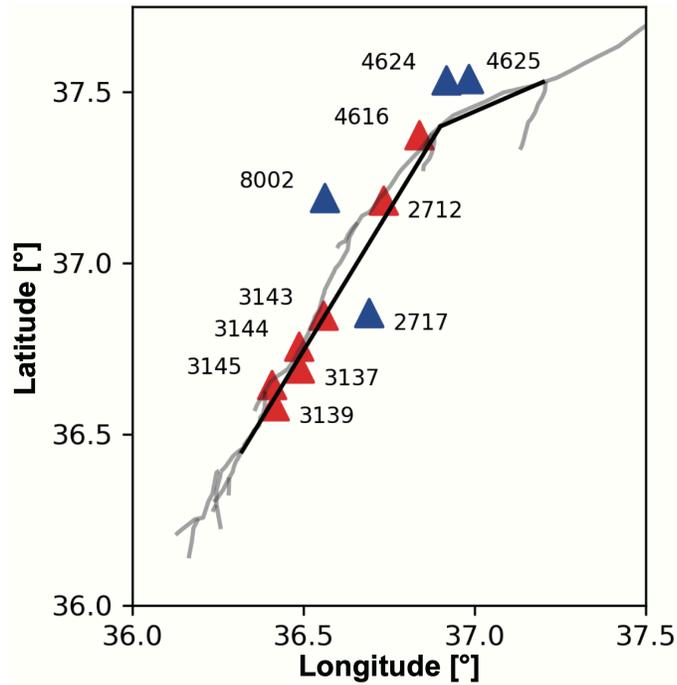

*Figure S1: Location of selected stations close to the fault surface rupture. We focus on the southern portion of the EAF, SW from its junction with the initial splay fault. The grey lines show surface ruptures (Reitman et al., 2023), the black line shows the simplified geometry considered to compute along-strike positions in Figure S3, and triangles show the position of stations at a distance from the fault shorter (red) and larger (blue) than 1 km.*

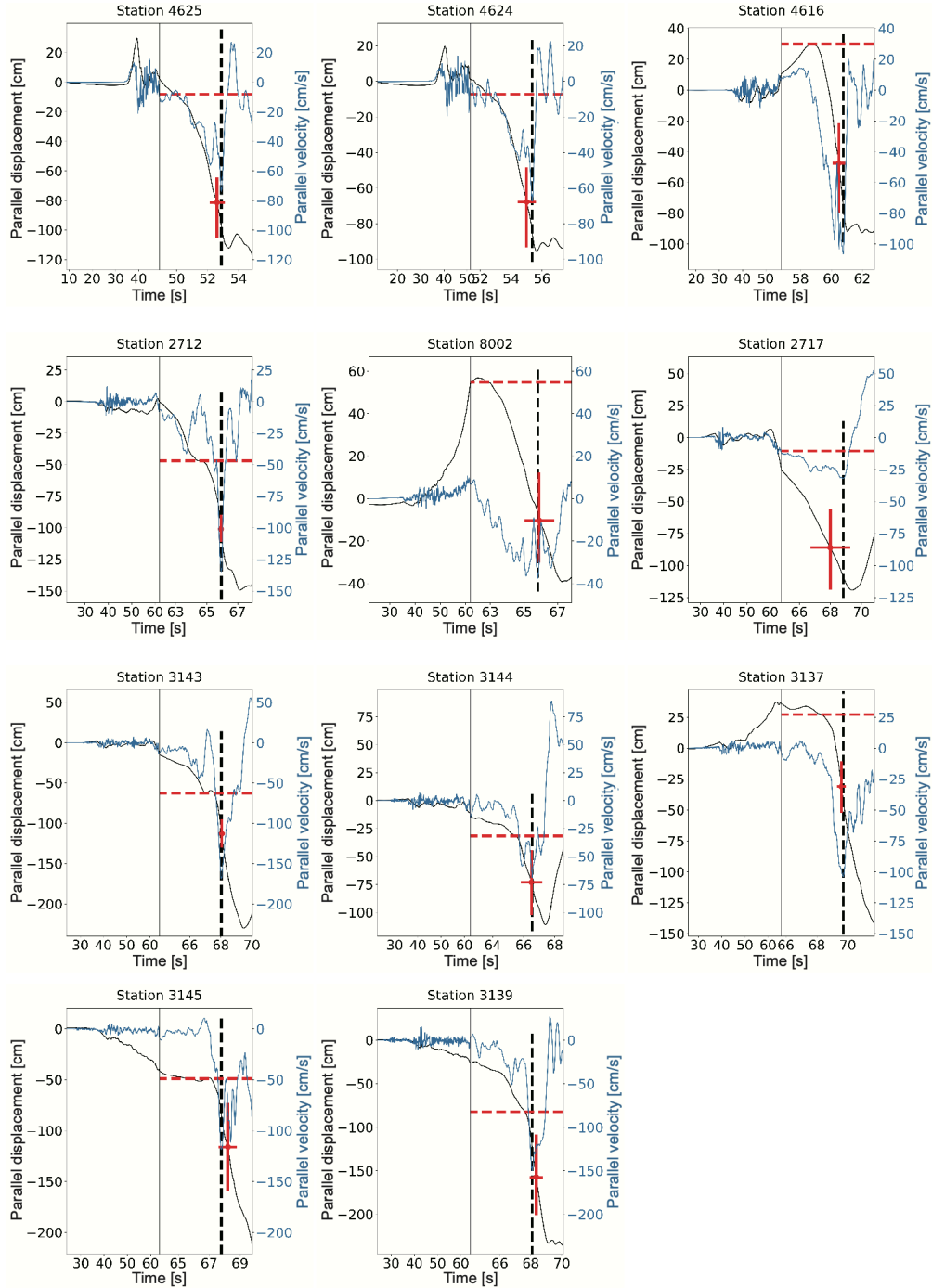

*Figure S2. Fault-parallel velocity (blue) and displacement (black) obtained by integrating acceleration data with baseline correction. The horizontal red-dashed line denotes the displacement level right before the passage of the SW-ward rupture front near the station; this value is taken as the reference to estimate $D_c''$. The vertical black dashed line indicates the time of maximum velocity. The red cross shows an*

*estimate of the uncertainty of maximum velocity (horizontally) and corresponding displacement (vertically). The red dot in their intersection indicates a mean value of peak-velocity time and displacement.*

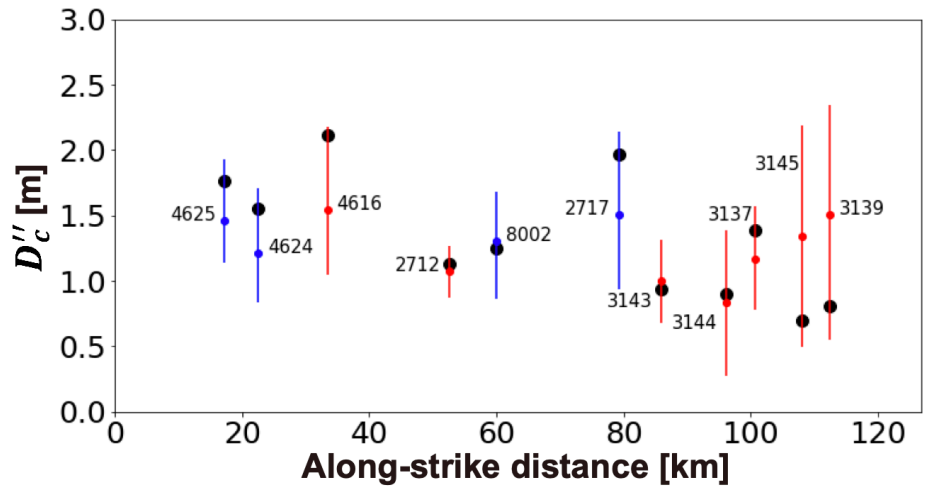

*Figure S3. Along-strike distribution of estimated $D_c''$. The distance in the horizontal axis is relative to the junction between the splay and main faults (see Figure S1). Large black dots show the value of $D_c''$ estimated from the displacement at the time of maximum velocity. Vertical bars show the $D_c''$ uncertainty inferred from the maximum velocity uncertainty, while small dots (in blue or red) show $D_c''$ at the center of the uncertainty interval. Colors indicate stations at a distance from the fault shorter (red) and larger (blue) than 1 km, as in Figure S1.*

## S2. Additional results for delayed rupture triggering on the SW segment of the main fault

In addition to the subshear case shown in the main text (Figures 5 and 6), we report here a case of supershear rupture along the splay fault, which successfully triggers first the NE segment of the main fault (Figures S4 and S5). The overall behavior of the triggered rupture along the main fault (Figure S5) is similar to that of the previous case (Figure 6). The main differences lie in the transient stress field before the splay-fault rupture is fully terminated at the junction. In the supershear case (Figure S4), the stress field comprises three parts that sequentially sweep along the SW segment of the main fault (Mello et al., 2010, 2016). The first one is carried by the dilatational field (zero curl) of the supershear front (Bhat et al., 2007b) and exerts a transient positive $\Delta CFS$ ($t_1$ to $t_2$ in Figure S4a, Figure S4b and c). The second one is carried by the S-wave Mach front and exerts a transient negative $\Delta CFS$ (around $t_3$ in Figure S4a, Figure S4d). The third one is carried by the trailing Rayleigh wave but is too weak to observe (Figure S4d). None of these three parts triggers a rupture along the SW segment of the main fault in the case shown in Figure S4; successful triggering along the SW segment occurs only after the NE segment is activated (Figure S4f and g), similar to the previous case shown in Figure 5.

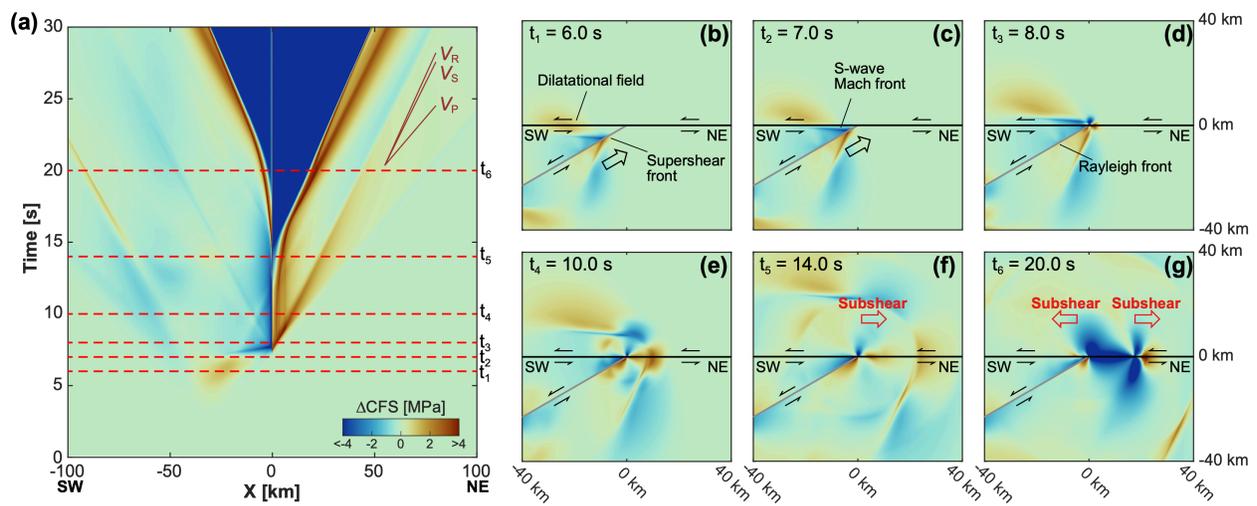

*Figure S4. Spatiotemporal distribution of Coulomb failure stress change ($\Delta CFS$) induced by a supershear rupture along the splay fault. (a) Evolution of $\Delta CFS$ projected along the main fault. Six times $t_1$ to $t_6$ are selected to highlight (1) when a dilatational stress lobe ($\Delta CFS > 0$) operates on the SW segment of the main fault, (2) when the S-wave Mach front ($\Delta CFS < 0$) is about to sweep over the SW segment next to the junction, (3) when the supershear rupture front just hits the junction along the splay fault, (4) when arrest waves start to radiate outward from the junction, (5) when rupture is just triggered along the NE segment of the main fault, (6) when the SW segment is activated by the stress transfer from the NE-ward propagating rupture. (b)-(g) Spatial distribution of $\Delta CFS$, resolved onto faults parallel to the main fault, at the six different times defined in (a). $f^{eff} = 0.48$ is assumed to compute $\Delta CFS$ (Eq. 3). Other model parameters are: $f_s^{sp} = 0.21$, $f_d^{sp} = 0.10$, $D_c^{sp} = 0.50\ m$; $f_s^m = 0.48$, $f_d^m = 0.29$, $D_c^m = 1.00\ m$.*

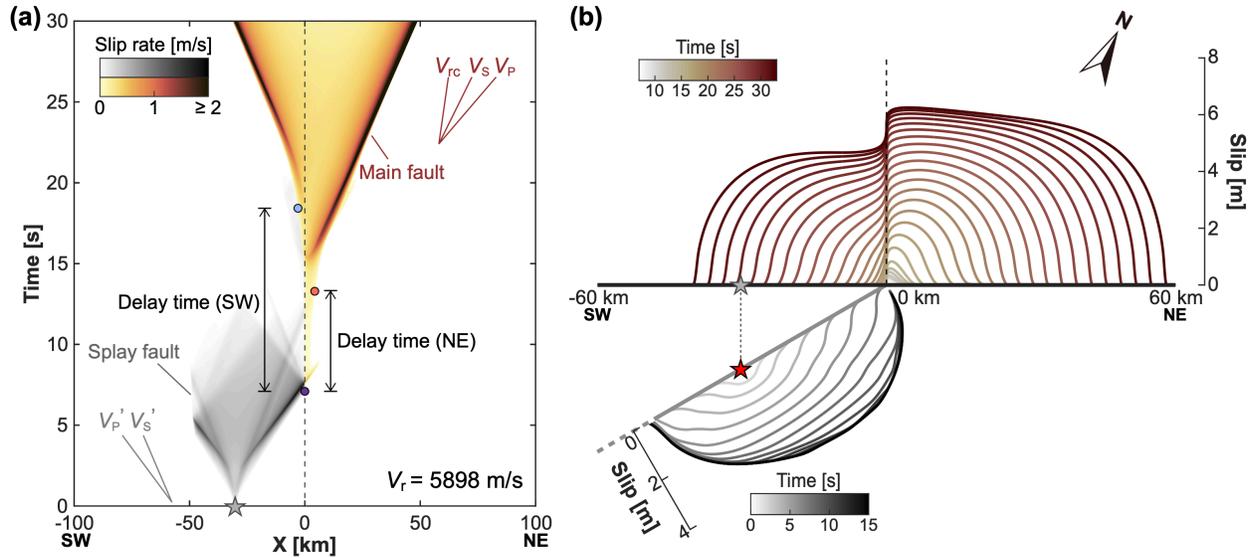

*Figure S5. Spatiotemporal distribution of (a) slip rate and (b) slip for the case shown in Figure S4. In (a), $V_r$ (5898 m/s) denotes the instantaneous propagation speed of the splay-fault rupture prior to its arrival at the junction with the main fault. The delay time is defined in the same way as in Figure 6.*

## S3. Additional results for early rupture triggering on the SW segment of the main fault

Here, we document the possibility of rupture triggering first on the SW segment of the main fault (Section 3.4) by reporting additional numerical simulation results. Figure S6 shows the evolution of ΔCFS for the case in Figure 11, in which a splay-fault subshear rupture triggers first the SW segment. The rupture along the SW segment is initiated around $t_2$ (Figure S6c) and shows bilateral propagation from $t_3$ to $t_4$ (Figure S6d and e). Earlier triggering can also be achieved by a supershear rupture along the splay fault (Figures S7 and S8). The dilatational stress carried by the splay-fault supershear front produces the earliest triggering along the SW segment (around $t_2$ in Figure S7). The following S-wave Mach front, though associated with a negative ΔCFS, does not stop the rupture triggered along the SW segment ($t_3$ to $t_4$ in Figure S7). Upon the arrival of the splay-fault rupture at the junction, a second rupture is triggered (after $t_4$ in Figure S7a), slightly skewed to the NE side of the junction along the main fault. Second rupture triggering can also be observed in the subshear case, initiated at the junction along the main fault (Figure S6), but is somewhat overshadowed by the first rupture. We expect this second rupture to follow the same mechanism as analyzed in Section 3.2.

Finally, we report an additional set of results on the delay time along the main fault (Figure S9). Here, we focus on the SW-ward rupture triggered along the main fault, including cases triggered first on the NE segment (marker with black edge color, as in Figure 6) and first on the SW segment (marker with red edge color, as in Figure 11). We do not consider second rupture triggering, even if it contains a SW-ward component (e.g., initiated at the fault junction in Figures S7 and S8). We overall find that, to trigger first the SW segment, the main fault must be initially close to failure (extremely low values of $S^m$).

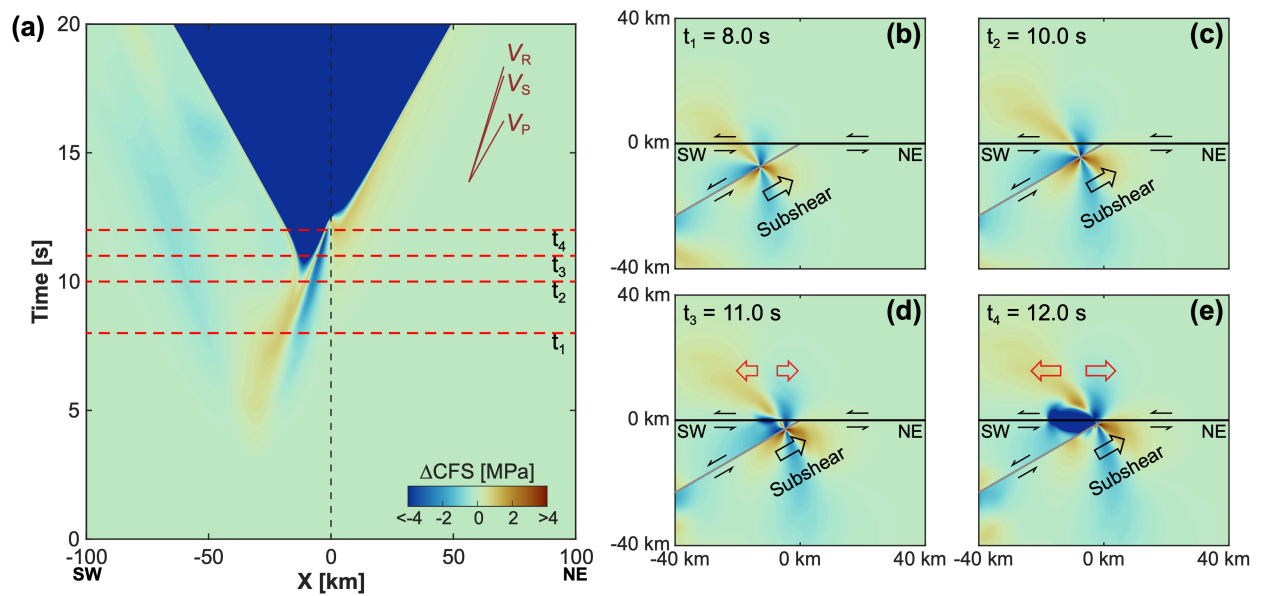

Figure S6. Similar to Figure S4 but for a subshear rupture along the splay fault that triggers rupture of the main fault first along its SW segment, before the splay-fault rupture arrives at the junction. $f^{eff} = 0.42$ is assumed to compute $\Delta CFS$ (Eq. 3). Other model parameters can be found in Figure 11, which shows the corresponding evolutions of slip rate and slip.

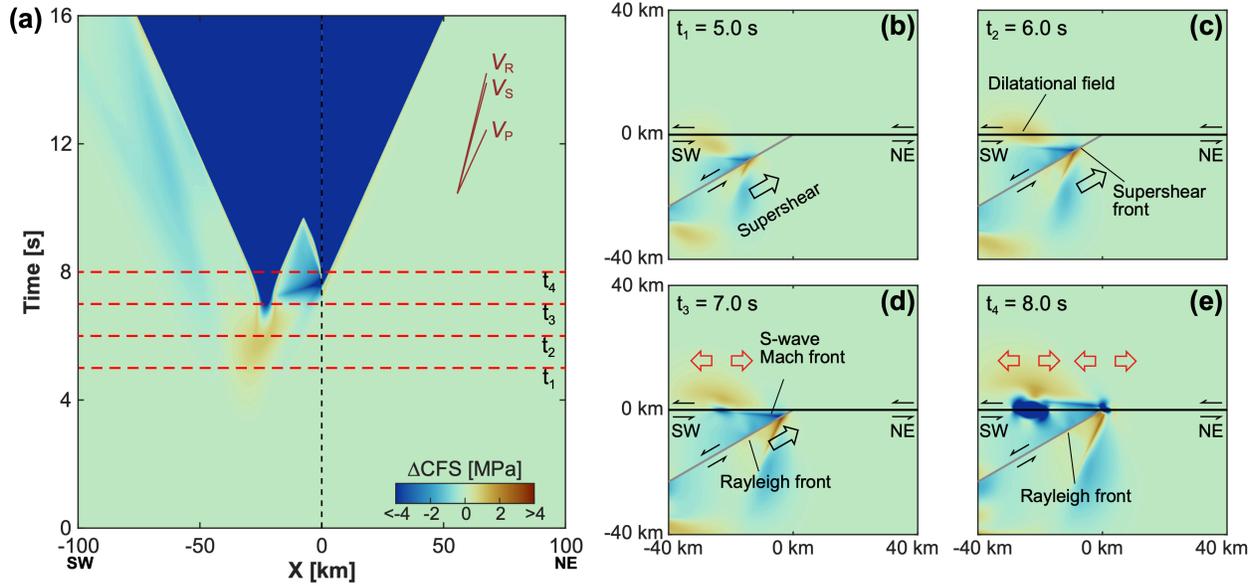

*Figure S7. Similar to Figure S6 but for a supershear rupture along the splay fault. Here also, rupture of the main fault is triggered first along its SW segment, before the splay-fault rupture arrives at the junction, and $f^{eff} = 0.42$. Other model parameters are: $f_s^{sp} = 0.21$, $f_d^{sp} = 0.10$, $D_c^{sp} = 0.50\ m$; $f_s^m = 0.42$, $f_d^m = 0.20$, $D_c^m = 0.50\ m$.*

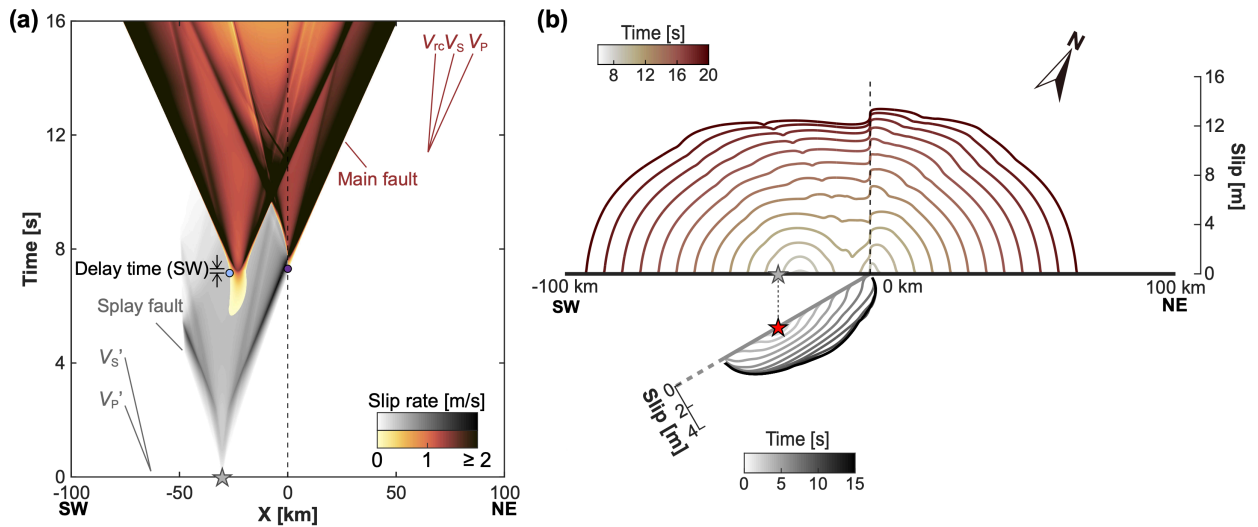

*Figure S8. Spatiotemporal distribution of (a) slip rate and (b) slip for the case shown in Figure S7.*

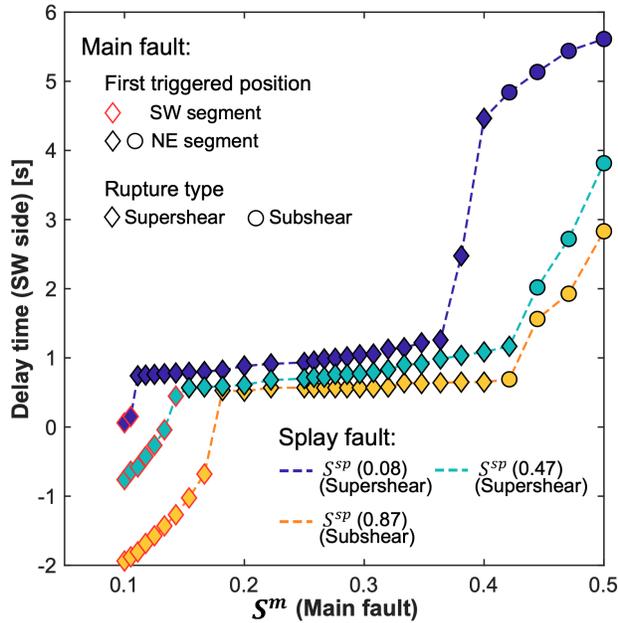

*Figure S9. Delay time along the SW segment of the main fault as a function of the seismic S ratio along the main fault ($S^m$) and along the splay fault ($S^{sp}$, indicated by curve and symbol-fill-in colors). Symbols with black and red edges correspond, respectively, to the cases with triggering first along the NE segment and SW segment of the main fault. Earlier triggering on the SW segment (negative or small positive delay time) occurs only at extremely low values of $S^m$, i.e. when the main fault is initially very close to failure. For the main fault, we vary $f_d^m$ (under fixed $f_s^m = 0.42$) to obtain different values of $S^m$. For the splay fault, we vary $f_s^{sp}$ (under fixed $f_d^{sp} = 0.10$) to obtain different values of $S^{sp}$. For both the main and splay faults, $D_c$ is fixed at 0.50 m.*

**References for Supplementary Material**